\documentclass[a4paper]{article}

\usepackage[ruled,linesnumbered,algo2e]{algorithm2e}
\usepackage{amsfonts}
\usepackage{amsmath}
\usepackage{array}
\usepackage{booktabs}
\usepackage{caption}
\usepackage{color}
\usepackage{etoolbox}
\usepackage{framed}
\usepackage{gensymb}
\usepackage{graphicx}
\usepackage{longtable}
\usepackage{multirow}
\usepackage{soul}
\usepackage{subcaption}
\usepackage{tikz}
\usepackage{todonotes}
\usepackage{units}
\usepackage{url}

\usetikzlibrary{arrows,decorations.pathreplacing,patterns}

\SetCommentSty{mycommfont}

\newcolumntype{P}[1]{>{\raggedright\hspace{0pt}\arraybackslash}p{#1}}

\newcommand{\mAa}{M-E1}
\newcommand{\mAb}{M-E2}
\newcommand{\mB}{M-BP}
\newcommand{\cons}{\ensuremath{\textsf{Cons}}}
\newcommand{\dir}{\ensuremath{\textsf{Direction}}}
\newcommand{\bp}{\ensuremath{\textsf{BehavioralProfiles}}}

\newcommand{\north}{\ensuremath{\textsf{North}}}
\newcommand{\south}{\ensuremath{\textsf{South}}}
\newcommand{\east}{\ensuremath{\textsf{East}}}
\newcommand{\west}{\ensuremath{\textsf{West}}}

\title{Detection and Quantification of Flow Consistency in Business Process Models}
\author{
	Andrea Burattin \thanks{A. Burattin (\emph{andrea.burattin@uibk.ac.at}), M. Neurauter and B. Weber are with the University of Innsbruck, Austria. V. Bernstein and P. Soffer are with the University of Haifa, Israel.} \and
	Vered Bernstein \and
	Manuel Neurauter \and
	Pnina Soffer \and
	Barbara Weber}

\begin{document}
\maketitle

\begin{abstract}
Business process models abstract complex business processes by representing them as graphical models. Their layout, solely determined by the modeler, affects their understandability. To support the construction of understandable models it would be beneficial to systematically study this effect. However, this requires a basic set of measurable key visual features, depicting the layout properties that are meaningful to the human user. The aim of this research is thus twofold. First, to empirically identify key visual features of business process models which are perceived as meaningful to the user. Second, to show how such features can be quantified into computational metrics, which are applicable to business process models. We focus on one particular feature, consistency of flow direction, and show the challenges that arise when transforming it into a precise metric. We propose three different metrics addressing these challenges, each following a different view of flow consistency. We then report the results of an empirical evaluation, which indicates which metric is more effective in predicting the human perception of this feature. Moreover, two other automatic evaluations describing the performance and the computational capabilities of our metrics are reported as well. \\[1em]
\end{abstract}

\section{Introduction}
\label{sec:introduction}

Business process modeling is a broad and important area for practice and for applied research. Process modeling refers to the representation of organizational or business processes in a graphical manner, usually as a flow of activities \cite{old6}. It is common across industries -- important for designing and improving business processes, analyzing industry goals and outcomes -- including organizational efficiency, revenues, and social impact \cite{old6}. For these purposes, the quality of the process model is of importance. Model quality has been classified to syntactic (``correctness'' of a model), semantic (the extent to which the model captures the behavior of the domain) and pragmatic (usefulness) quality \cite{old9}. When focusing on pragmatic quality, one of the important aspects considered is the understandability of the model by a human user. Indeed, efforts have been made, attempting to study and to improve user comprehension of process models \cite{old14,old24}.   

A large body of research has addressed factors that influence model understandability, relating both to business process models and to other kinds of conceptual models. Much attention in this respect has been given to semantic clarity of the modeling language \cite{old15,old22} and to the graphical elements of the modeling language \cite{old17,old23}. Additional factors identified relate to properties of an individual model (e.g., complexity metrics \cite{old3,old12,old29,old30}). Visual features of elements in a model \cite{old25,old27} have been studied, specifically the effect of what is sometimes called ``secondary notation'' on model understandability \cite{old10,old26}. In contrast, the specific layout of a model has received little attention. To the best of our knowledge, only a few studies have investigated how layout features of a process model affect its understandability. One specific example is~\cite{Figl2015}, addressing the flow direction of a process model and its effect on model understanding. This investigation addressed only consistent flow directions (e.g., left to right, top to bottom), not addressing the possibility of change in the direction of the model. 

Cognitive psychology research has shown that the appearance of a model in general has a significant effect on user comprehension \cite{old11,old17}. Thus, the visual layout of a process model is central to achieving its aims -- effectively communicating the intended process, ensuring comprehension by its users, and enabling revision and improvement of the process model. Yet, we currently lack an agreed upon set of concepts for describing and characterizing layout properties as perceived by humans. 

In the context of modeling and model generating tools, model layout has received attention and has been characterized by precisely defined properties. These tools typically include functionality that determines the layout and rearranges the elements in the model in order to improve its readability (e.g., \cite{old4,old5,old7}). The algorithms that are employed relate to precisely defined  properties, such as avoiding line crossings in the model, alignment of model elements, usage of straight angles with the goal to produce a ``neatly'' appearing model. However, there is no indication regarding how comprehensively these properties correspond to and capture the human perception of the model. In fact, there is no cognitive anchoring of the specific selection of features that are currently addressed. 

We believe that a set of concepts describing layout features should comprehensively correspond to how humans perceive the layout of the model. At the same time, it should be precisely defined and allow quantification and measurement of layout properties, serving several purposes. First, it will permit a more focused study of the effect of process model layout on model understanding. Such studies may address specific properties or combinations of properties. Second, it can become a basis for guiding the creation of process models. Currently, although there have been some broad efforts to guide modeling from a visual perspective -- such as 7PMG \cite{old13} -- process modelers individually decide how to design the process model layout. Note that most of the modeling guidance that is available (e.g., in 7PMG) is semantic and structural, while the visual perspective is only addressed to a very limited extent.  Third, systematically developed layouting guidelines may support the training of modelers. When applied in an online setting, they can further serve as basis for intelligent modeling environments that provide feedback to modelers (on how to improve the model). Finally, they can serve for the development of automatic layout features of modeling tools. 

The aim of this paper is to take a step towards a set of human-meaningful and measurable layout features of process models. We start by qualitatively identifying key visual layout features in an exploratory study based on human perceptions. We then focus on the specific feature of consistency of flow direction to illustrate the challenges of transforming an intuitively perceived feature into a precise metric. We propose two different approaches operationalized into three metrics that address these challenges, each following a different view of flow consistency. Finally, we report an empirical evaluation, which indicates the extent to which these metrics are more consistent with the human perception of this feature. Two automatic evaluations, with the aim of providing a better characterization of our metrics are reported as well.

Accordingly, the remainder of the paper is structured as follows: Section~\ref{sec:study} presents the methodology and setting of the exploratory study; Section~\ref{sec:analysis} reports the findings of this study as a list of identified layout features; Section~\ref{sec:quantification} focuses on the consistency of flow feature and presents alternative approaches for its quantification.
Section~\ref{sec:evaluation} deals with the empirical evaluation of the proposed approaches and discusses the findings, while Section~\ref{sec:relatedwork} reviews related work. Section~\ref{sec:conclusions} provides the conclusions and highlights implications of this study for future research.

\section{Exploratory Study}
\label{sec:study}

The exploratory study was guided by the research question of what layout features of process models are perceived as meaningful by model users. Due to the nature of the question, which seeks to discover features rather than to corroborate hypotheses, the study was exploratory in nature. To identify candidate visual features of a process model, an empirical qualitative study was conducted. Qualitative data gathering was needed in order to get an understanding of a user's point of view \cite{old1}. Acquiring knowledge from participants was essential to understand how they perceive the visual layout of business process models.

\subsection{Setting}

The exploratory study took place at the Department of Information Systems at the University of Haifa. Participants in the study were 15 undergraduate and 7 graduate students. Participants all had some knowledge of business processes modeling. All participants came with similar educational background -- all took a variety of information systems analysis courses and studied modeling languages. Participation in the study was voluntary. 

The study included questionnaires and interviews. First, 15 undergraduate students were asked to fill out a questionnaires. Following an initial screening of the answers that were obtained, additional 7 graduate students were interviewed to gain a deeper understanding. The interviews were based on the questionnaires, but allowed for interaction and prompting deeper explanations. Thus, interviews were used in order to get a better understanding of the user perspective on the visual layout of business process models \cite{old18}.

\subsection{Data Collection Process}

\begin{figure}
	\begin{subfigure}[b]{.49\textwidth}
		\includegraphics[width=\textwidth]{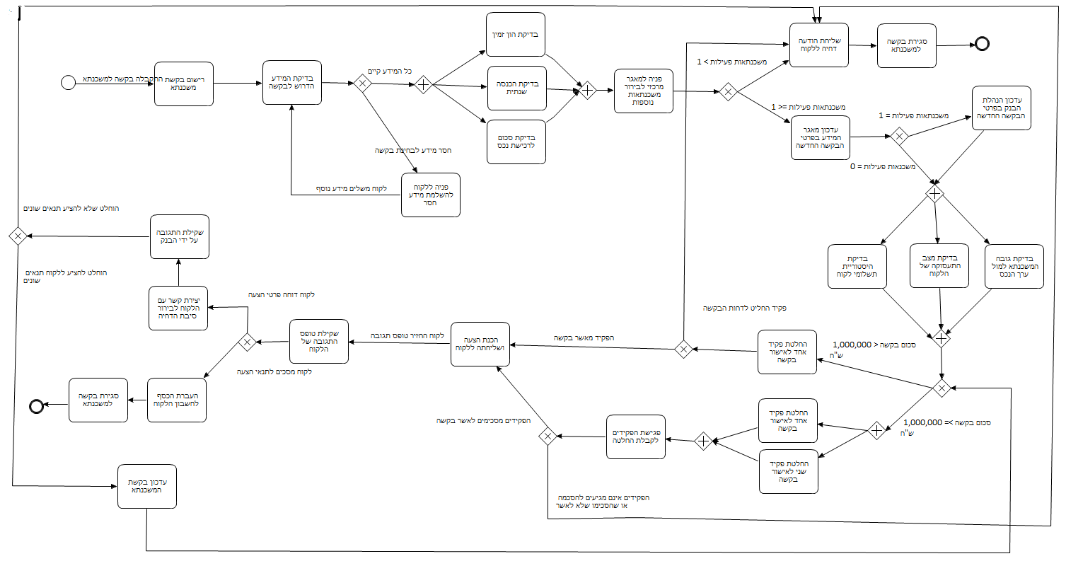}
		\caption{}
	\end{subfigure}\hfill
	\begin{subfigure}[b]{.49\textwidth}
		\includegraphics[width=\textwidth]{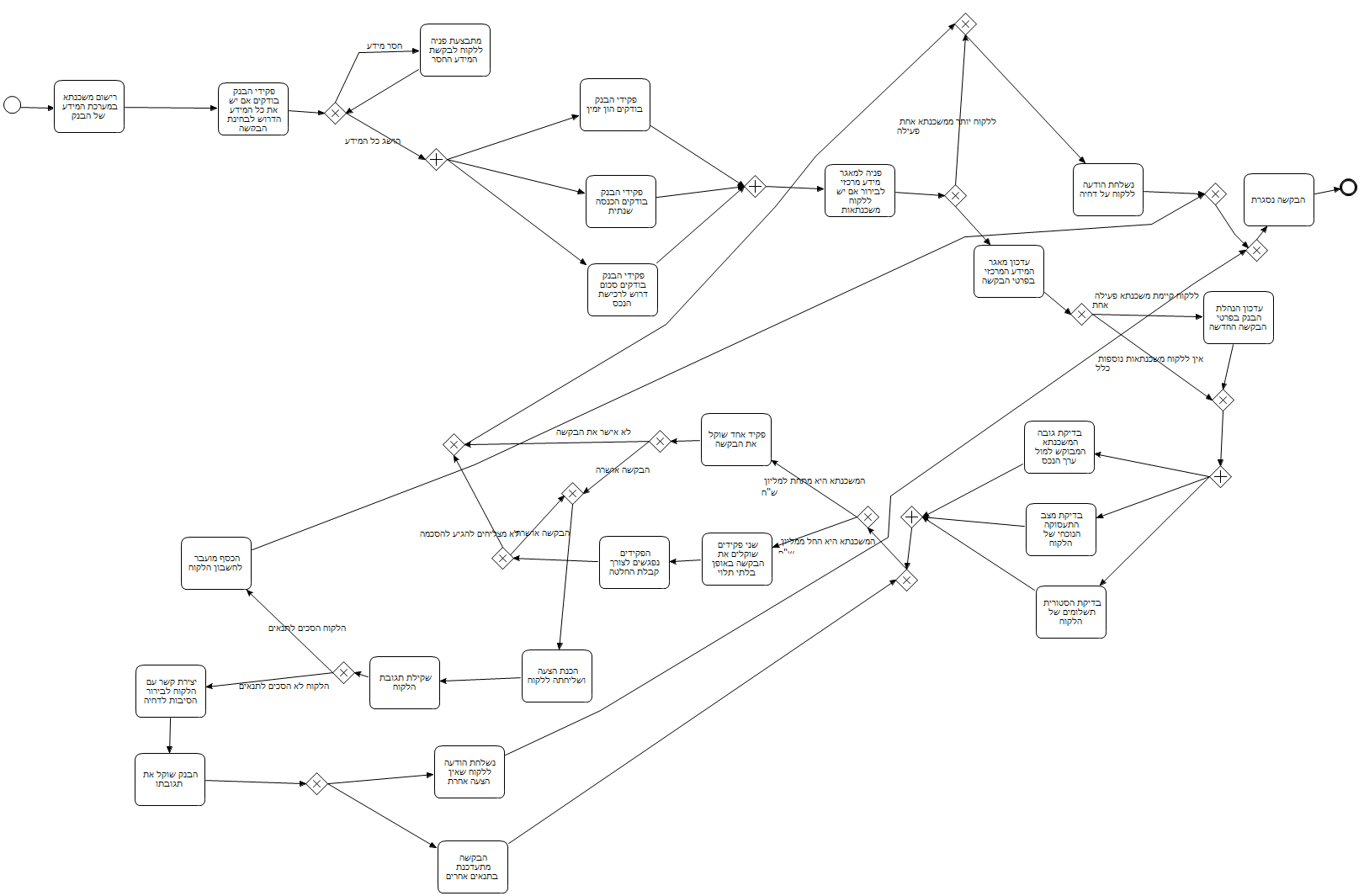}
		\caption{}
	\end{subfigure}
	\caption{Example pair of models from the questionnaire.}
	\label{fig:questionnaire-example}
\end{figure}

\paragraph{Questionnaires.}
The goal of the questionnaire\footnote{All the material used for this study is available at \url{http://bpm.q-e.at/experiment_flow_consistency}.} was to understand participants' beliefs and perceptions of the visual layout features of process models. A pilot questionnaire was given ahead of time to three participants in order to simplify and improve the questions in the questionnaire. The questionnaire had 5 pairs of BPMN models. The models were made small to fit a single page, yet it was ensured that their structure was clearly visible. All activity labels or edge labels were blurred in order to have participants address the visual aspect of the model exclusively and not ``read into it''. Some of the pairs included two models which appeared to be visually different according to the judgment of one of the researchers, while others appeared visually similar. In other words, the pairs were selected to have different levels of similarity according to the researcher's judgment. The models were all presented in black and white in order to have participants focus on layout features of the models. Color might have drawn much attention and blur the effect of less dominant features. An example pair from the questionnaire is shown in Fig.~\ref{fig:questionnaire-example}. The questionnaire consisted of the same set of questions referring to each of the pairs. This set included one question asking the participants to rate the visual similarity of the models on a 7-point Likert scale. The goal of using a Likert scale was to prompt participants to actually look at the figures, compare their layout, and evaluate their similarity. Following the Likert scale evaluation, two open-ended questions were presented to the participants, asking them to indicate differences and similarities between the models (at least three of each). Only the answers of the open-ended questions were considered in the data analysis; the Likert-scale evaluation was only intended to prompt the comparison of the models and was not analyzed afterwards.

\paragraph{Interviews.}
The interviews were semi-structured, based on the questionnaires. First, participants were asked to complete the questionnaire. Next, the participants were asked specific questions about their answers, prompting additional explanations about specific differences or similarities between the models of each pair. In particular, the questions related to features that support or hamper the understandability of the models in order to encourage the participants to engage in the specific appearance features. Participants were asked questions such as: \emph{At a first glance, which model seems easily readable and why? Which model is less understandable? What visual features would you change here to make it easier to understand?} The participants were also asked to point out differences between models and indicate their preferences in regards to comprehension of the models.

\section{Analysis and Findings}
\label{sec:analysis}

The data collected from the questionnaires and interviews was qualitatively coded and classified into categories based on data collected from participants.

\subsection{Analysis}

The analysis considered the text of the written questionnaires and the interviews text. Using qualitative data analysis methods \cite{old28}, textual segments were coded by the model(s) they referred to and classified to categories of features, which were later on aggregated to higher-level categories.  Table~\ref{tbl:data-analysis} describes the steps taken during the data analysis phase. Saturation of the categories was reached by the fourth interview, so eventually no new categories were found with additional data analysis.

\begin{table}
	\centering
	\begin{tabular}{l|P{25em}}
		\toprule
		Stage I & Getting familiar with the data -- reading and interpreting the data collected using the questionnaires; collecting additional data through interviews \\
		\midrule
		Stage II & Identifying repeating elements -- color categorizing concepts which repeat in the different questionnaires and interviews. Saturation of categories was reached by the 4\textsuperscript{th} interview \\
		\midrule
		Stage III & Recognizing super categories -- finding higher level categories grouping lower-level ones \\
		\bottomrule
	\end{tabular}
	\caption{Data Analysis Process}
	\label{tbl:data-analysis}
\end{table}

\subsection{Categories}

The next stage was to identify and define categories of visual features in the models. Table~\ref{tbl:categories} summarizes the categories found by qualitative analysis. It provides a definition for each category, and examples of supporting statements taken from the questionnaires and interviews. The lower-level categories are grouped under higher-level ones (where applicable).

{\small
\begin{longtable}{P{.2\textwidth}|P{.4\textwidth}|P{.3\textwidth}}
\caption{Data Analysis Process\label{tbl:categories}} \\
\toprule
\textbf{Feature} & \textbf{Description} & \textbf{Example Reference from Data} \\
\midrule
\endfirsthead
\caption{Data Analysis Process} \\
\toprule
\textbf{Feature} & \textbf{Description} & \textbf{Example Reference from Data} \\
\midrule
\endhead
\bottomrule
\endlastfoot
\multicolumn{3}{l}{\textbf{Edges}} \\
\midrule
Length of edges &
	The length of the edges in the model. A model may vary consisting very short edges (creating a dense model) to very long edges (creating a widely spread model), or a mixture of lengths &
	``\emph{The model on the right doesn't seem right since there are many long edges throughout the model}'' \\
	\midrule
Edges style: straight, curved or with bending points &
	Edges can be straight, or curved or they may consist of one or more bending points, which  divides the edge  into two segments or more &
	``\emph{Need to straighten all the broken edges}'' \\
	\midrule
Crossing edges &
	Edges that cross each other -- intersect with other edges. Intersecting edges might create confusion when following the flow of the model &
	``\emph{There are edges here that just go one on top of the other}'', ``\emph{This looks like a spider web}'' \\
	\midrule
Text on edges &
	Existence and amount of text annotations on edges. The text can either be descriptive or conditional &
	``\emph{When something is written on the edge, it is difficult to understand which edge it refers to}'' \\
	\midrule
Number of ending points &
	The total number of ending points in the model. An ending point is an end event or an element with no outgoing edges &
	``\emph{There are many ending points}'', ``\emph{One ending point connected to many edges, appears like a loop}'' \\
	\midrule
Angles &
	The angles used in bending points of edges: 90\degree{} angles, 180\degree{} angles, angles larger than 45\degree{}, angles smaller than 45\degree{} &
	``\emph{I would improve the angles in this model to be 90\degree{} angles}'', ``\emph{Change the edges to be straight lines}'' \\

\midrule
\multicolumn{3}{l}{\textbf{Model's Structure}} \\
\midrule
Model's shape &
	The general shape of the model refers to the way the model is spread on the canvas. This usually is characterized as a square or rectangle &
	``\emph{The structure in both models is horizontal}'' \\
	\midrule
Model's area &
	The area taken by the model on the canvas &
	``\emph{The size of the models is different}'' \\

\midrule
\pagebreak
\multicolumn{3}{l}{\textbf{Model's Direction}} \\
\midrule
General direction &
	The general direction/flow of the model. The direction of the model can be characterized as vertical or horizontal: Left-right direction, right-left direction, top-bottom direction, bottom-top direction &
	``\emph{This model goes in a clear direction}'', ``\emph{Both models are vertical}'' \\
	\midrule
Placement of ending event &
	The location of ending points in the model in relation to the starting point of the model &
	``\emph{Location of the ending point makes it clear where the process ends}'' \\
	\midrule
Branching off &
	Branching off of the model from one main path to more than one, where each branch flows in a different direction &
	``\emph{I don't like to wonder where an edge leads to}'' \\
	\midrule
Consistency of flow &
	The flow of the model can be in one definite direction from the beginning till the end of the model. Alternatively, it can be unclear or changing throughout the model to different directions & 
	``\emph{There is a change in the direction of the model}'', ``\emph{Both models are built stepwise}'' \\
	\midrule
Symmetry in blocks &
	Referring to structured blocks in the model -- symmetry of elements arrangement across the block &
	``\emph{This block in the model is very symmetrical and therefore very understandable}'' \\
	\midrule
Alignment in the model &
	Alignment of the elements in the model in relation to each other &
	``\emph{This model is clearer because of the alignment of the whole model. It is very aesthetic}'' \\
\end{longtable}}
\section{Quantification of Flow Consistency}
\label{sec:quantification}

In Table~\ref{tbl:categories} the different visual features identified in the exploratory study are highlighted. In order to enable the systematic investigation of how these features impact understandability it is necessary to transform them into metrics that can be automatically computed. In our previous paper~\cite{Bernstein2015}, we already reported details about some of these metrics. In the following we will focus on one of the identified features, namely, \emph{consistency of flow}. In general, we can state that the consistency of flow measures the extent to which the layout of a process model reflects the temporal logical ordering of the process. This metric, not analyzed in details in our previous work, is particularly challenging since it involves ``high-level concepts'' and how such concepts are represented (e.g., the ``layout'' of a process model is \emph{realized} by its nodes and edges). Moreover, there are several different ways of computing it, and it is not obvious which approach would most closely reflect human perception.

\begin{figure}
	\begin{subfigure}[b]{\textwidth}
		\includegraphics[width=\textwidth]{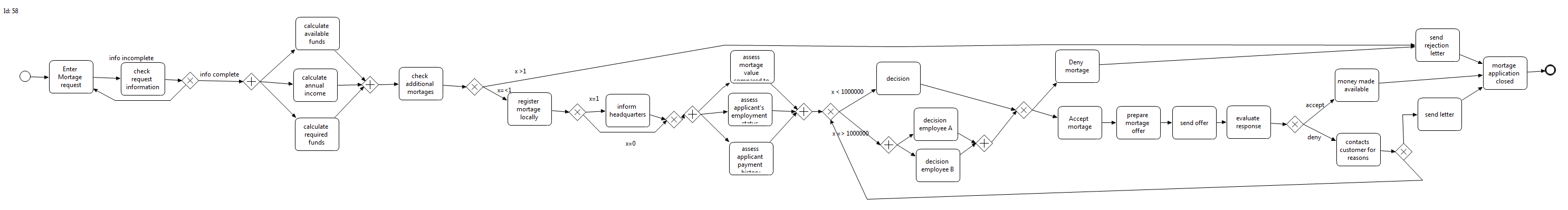}
		\caption{Process model with a consistent direction of the flow.}
		\label{fig:models-examples:linear}
	\end{subfigure}\\[1em]
	\begin{subfigure}[b]{.55\textwidth}
		\includegraphics[width=\textwidth]{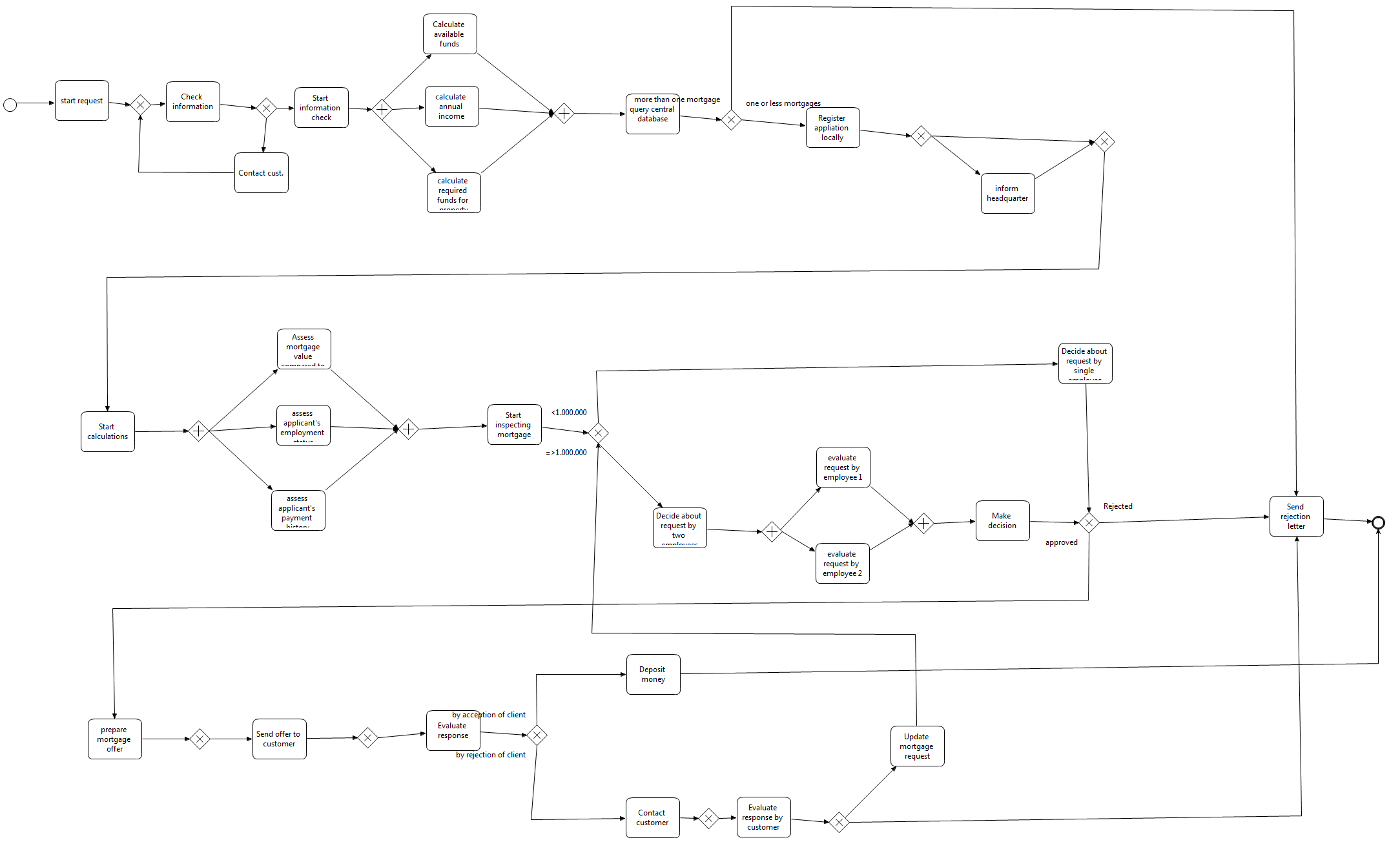}
		\caption{Model with some violations of the flow consistency.}
		\label{fig:models-examples:fragment}
	\end{subfigure}\hfill
	\begin{subfigure}[b]{.4\textwidth}
		\centering
		\includegraphics[width=\textwidth]{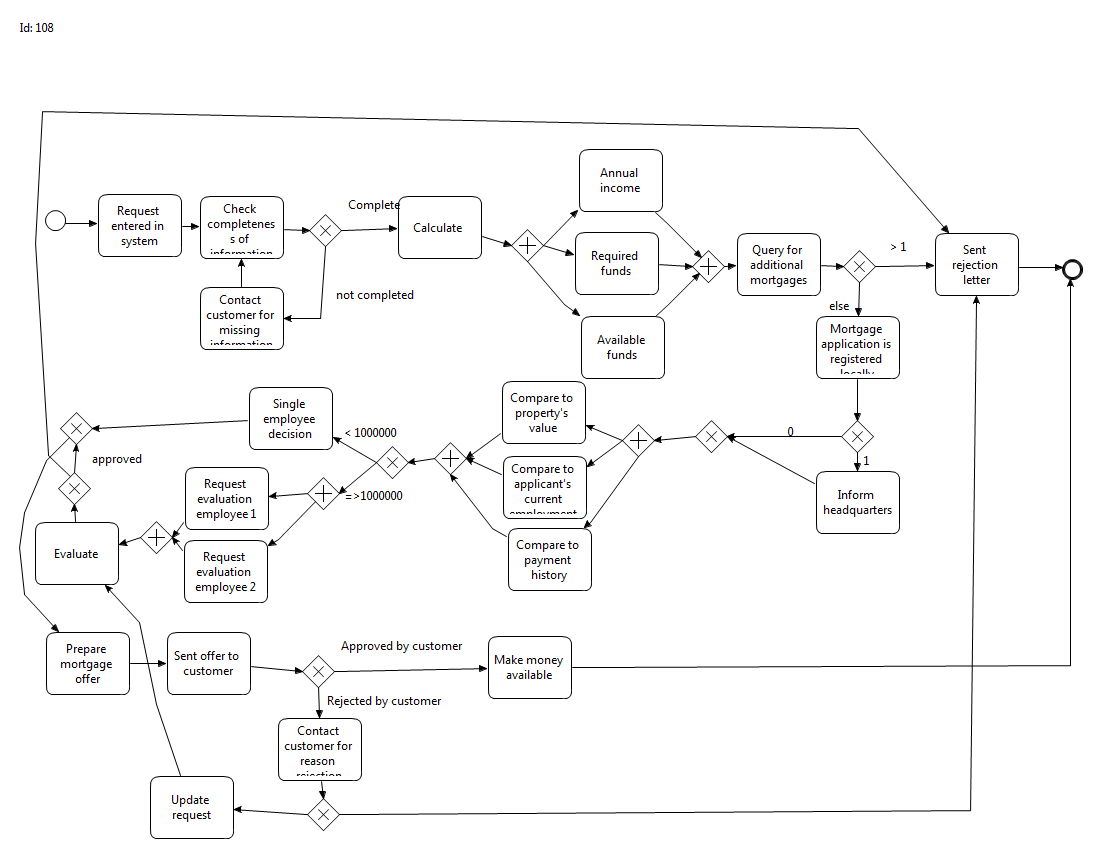}
		\caption{Model without a strong flow consistency.}
		\label{fig:models-examples:messy}
	\end{subfigure}
	\caption{Examples of models with different layouts, obtained starting from the same process description.}
	\label{fig:models-examples}
\end{figure}

For example, let us consider the models in Figure~\ref{fig:models-examples}. Fig.~\ref{fig:models-examples:linear} depicts a model which structures the 
flow in one, horizontal, left-to-right direction. The model in Fig.~\ref{fig:models-examples:fragment}, in turn, contains three ``horizontal lines'', each of them with a clear direction. In this case, in order to read the complete process, it is necessary to change the reading direction between each line: the reader has to ``go back'' with the eyes to the left side before continuing to the right again. Therefore, the flow direction of this process is less consistent compared to Fig.~\ref{fig:models-examples:linear}. For the model in Fig.~\ref{fig:models-examples:messy} it is even more difficult to identify a clear flow direction.

For most of the visual features described in Table~\ref{tbl:categories}, a clear mathematical quantification is mentioned in our previous work~\cite{Bernstein2015}. The description of the consistency of the flow, however, is just sketched.
In particular, several options exist on how to approach the quantification. Metrics for the automatic identification of changes in the flow direction can be based on global or local features. Global features look at the overall shape of the model and allow to detect the flow consistency based on the ``global shape'' of the process. Local features, in turn, consider how activities (i.e., vertices of the graph representation of the process) and sequences (i.e., edges of the graph representation of the process) are positioned in relation to each other. Concerning global features, human beings can easily detect that the model illustrated in Fig~\ref{fig:models-examples:linear} consists of one horizontal line, while the model shown in Fig.~\ref{fig:models-examples:fragment} is spread over several lines. For algorithms, however, the identification of such patterns is rather difficult, while local features can be much better dealt with.
Since our goal is to provide algorithmic solutions we decided to follow the second approach, based on local features. To this extent, and since we also want to closely reflect the human perception, we devised three different metrics. The first two metrics calculate the direction of each edge; determine the most frequent direction, based on majority voting; and then, based on this predominant direction, compute the extent to which the edges of the model are consistent with this direction. The third metric, instead, looks at pairs of activities and determines whether their positioning reflects their temporal local ordering.

\subsection{Graphical Representation of Processes}

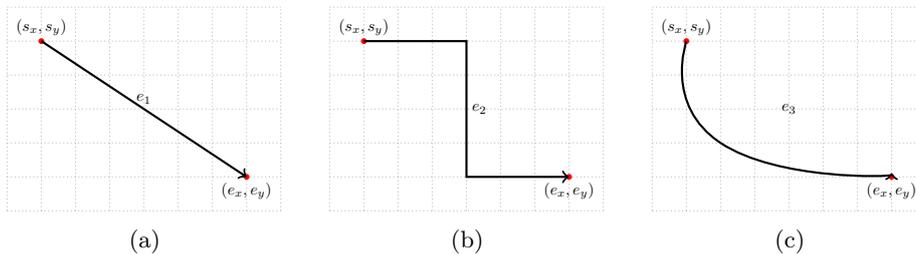
\begin{figure}
	\centering
	\begin{subfigure}[b]{.3\textwidth}
		\begin{tikzpicture}[scale=0.9, every node/.style={scale=0.6}]
		\draw[step=5mm, gray!50, style=densely dotted] (0,0) grid (4,3);
		\filldraw[red] (0.5,2.5) circle (1pt) node [above, black] {$(s_x, s_y)$};
		\filldraw[red] (3.5,0.5) circle (1pt) node [below, black] {$(e_x, e_y)$};
		\draw[thick, ->] (0.5,2.5)  -- (3.5,0.5);
		\draw (2,1.5) node [above] {$e_1$};
		\end{tikzpicture}
		\caption{}
	\end{subfigure}
	\hfill
	\begin{subfigure}[b]{.3\textwidth}
		\begin{tikzpicture}[scale=0.9, every node/.style={scale=0.6}]
		\draw[step=5mm, gray!50, style=densely dotted] (0,0) grid (4,3);
		\filldraw[red] (0.5,2.5) circle (1pt) node [above, black] {$(s_x, s_y)$};
		\filldraw[red] (3.5,0.5) circle (1pt) node [below, black] {$(e_x, e_y)$};
		\draw[thick, ->] (0.5,2.5) -- (2,2.5) -- (2,0.5) -- (3.5,0.5);
		\draw (2,1.5) node [right] {$e_2$};
		\end{tikzpicture}
		\caption{}
	\end{subfigure}
	\hfill
	\begin{subfigure}[b]{.3\textwidth}
		\begin{tikzpicture}[scale=0.9, every node/.style={scale=0.6}]
		\draw[step=5mm, gray!50, style=densely dotted] (0,0) grid (4,3);
		\filldraw[red] (0.5,2.5) circle (1pt) node [above, black] {$(s_x, s_y)$};
		\filldraw[red] (3.5,0.5) circle (1pt) node [below, black] {$(e_x, e_y)$};
		\draw [thick, ->] plot [smooth,tension=1] coordinates {(0.5,2.5) (1,1) (3.5,0.5)};
		\draw (2,1.5) node {$e_3$};
		\end{tikzpicture}
		\caption{}
	\end{subfigure}
	
	\caption{Three differently shaped edges $e_1$, $e_2$ and $e_3$, that, using our abstraction framework, are equally represented with its starting and ending coordinates $((s_x, s_y), (e_x, e_y))$.}
	\label{fig:edge-styles-abstraction}
\end{figure}

In order to present our metrics, we define the graphical representation of a process model as a diagram $G = (V, E, L_V, L_E)$. $G$ is a tuple composed of the set of vertices $V$ and the set of directed edges $E \subseteq V \times V$. Each vertex\footnote{In our definition, vertices represent all the graphical nodes that we might observe in a process model, not only activities. For example, considering BPMN processes, we should also consider events, gateways, etc.} and each edge have a corresponding graphical representation. Therefore, differently from typical \emph{graph} characterizations available in the literature \cite{Cormen2001}, we added two more relations $L_V$ and $L_E$, with information regarding the positioning of the respective elements on the modeling canvas (i.e., coordinates on the Cartesian plane). For vertices we consider the central point of its graphical representation $L_V : V \to (\mathbb{N} \times \mathbb{N})$. For edges, in turn, we consider two coordinates, one representing the starting and one the ending points. $L_E : E \to (\mathbb{N} \times \mathbb{N}) \times (\mathbb{N} \times \mathbb{N})$. Note that this edge representation abstracts from the actual path of the edge (cf. Figure~\ref{fig:edge-styles-abstraction}) and is therefore able to deal with edges whose layout is typical of business process models.
\begin{figure}
	\begin{subfigure}[b]{.3\textwidth}
		\centering
		\includegraphics[width=.7\textwidth]{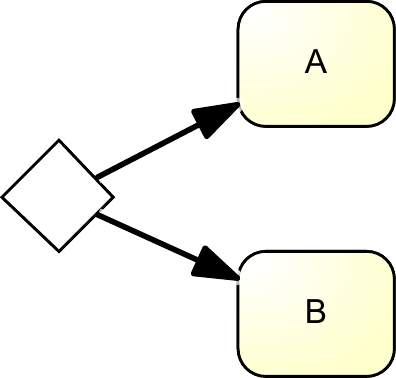}
		\caption{}
	\end{subfigure}
	\begin{subfigure}[b]{.3\textwidth}
		\centering
		\includegraphics[width=.7\textwidth]{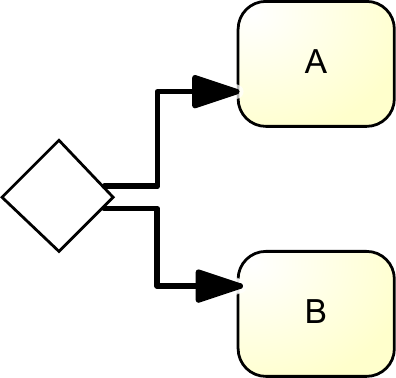}
		\caption{}
	\end{subfigure}
	\begin{subfigure}[b]{.3\textwidth}
		\centering
		\includegraphics[width=.7\textwidth]{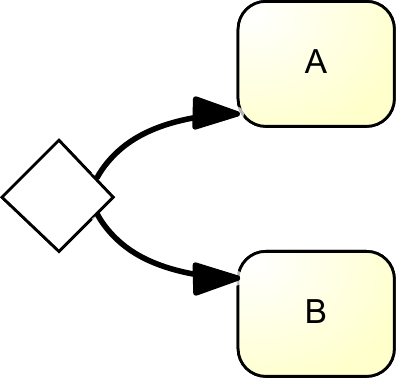}
		\caption{}
	\end{subfigure}
	\caption{Snippets of three common representations of edges outgoing from a BPMN gateway to activities. Edges of each snippet are laid out according to the edge shapes reported in Fig.~\ref{fig:edge-styles-abstraction}.}
	\label{fig:edge-styles-abstraction-processes}
\end{figure}
For example, Figure~\ref{fig:edge-styles-abstraction-processes} depicts three possible and common ways of representing edges exiting from a gateway. The three different representations are based on the edge styles reported in Fig.~\ref{fig:edge-styles-abstraction}, and can also be observed in the processes depicted in Fig.~\ref{fig:questionnaire-example} and~\ref{fig:models-examples}.

\subsection{Metrics Based on Edges' Directions}
\label{sec:edges-direction-metrics}

In this section we introduce a function, $\cons(G)$, which will be operationalized into two different metrics. Both metrics quantify the consistency of flow, i.e., the extent to which the layout of a process model reflects the temporal logical ordering of the process. To determine the extent of consistency, these metrics primarily focus on the edges of the process model and quantifies the consistency of all the edges $E$ in graph $G$.

\begin{figure}
	\centering
	\includegraphics[width=.7\textwidth]{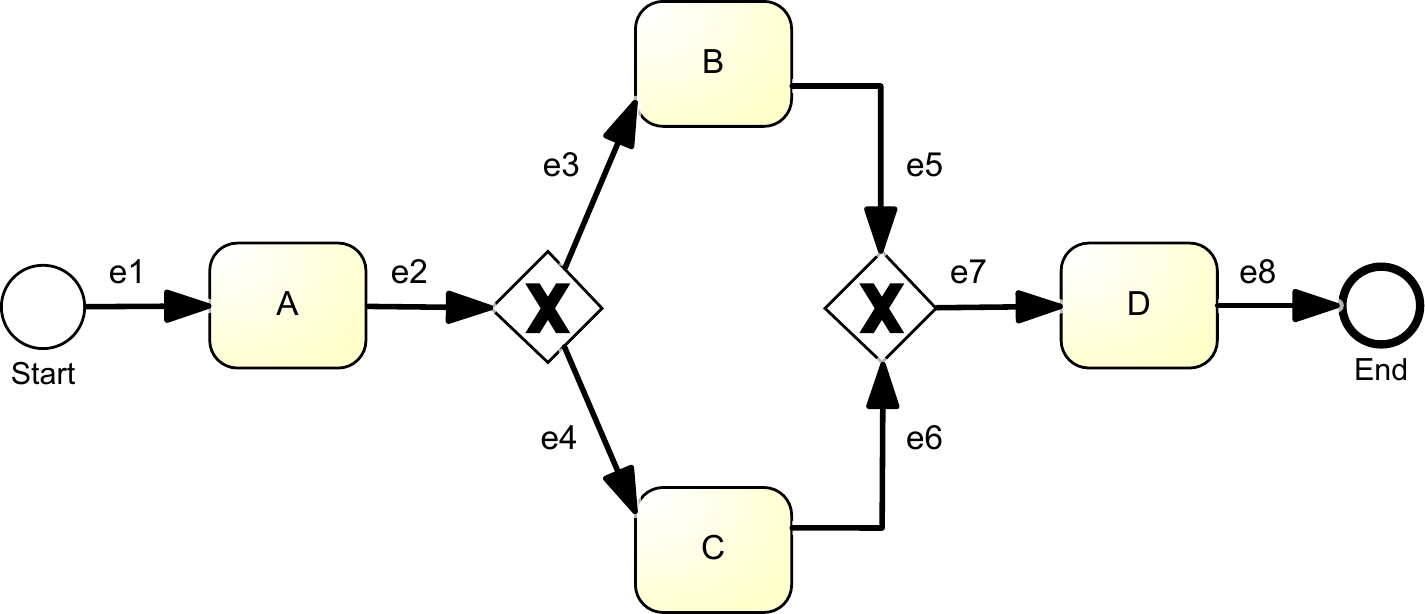}
	\caption{Illustrating example for metrics based on edges' direction.}
	\label{fig:motivating-process}
\end{figure}

To outline the idea behind these metrics, consider the process model depicted in Fig.~\ref{fig:motivating-process}, which shows a typical layout for a business process model. The fundamental idea is to determine, in a first step, the predominant flow direction of the process graph $G$. When looking at Fig.~\ref{fig:motivating-process} we can easily see that the diagram points east, i.e., the predominant flow direction is east. Then in a second step the metrics check the consistency of all the edges of graph $G$ with the predominant flow direction. 
Analyzing each edge, we can see that $e_1$, $e_2$, $e_7$ and $e_8$ clearly point east, i.e., they are consistent with the predominant flow direction. For edges $e_3$, $e_4$, $e_5$, and $e_6$ it is less obvious since these edges cannot be assigned to one clear direction easily. Following a na\"ive approach, we could just consider the most predominant direction of the edge (like we did for the entire process graph): we may classify edges $e_3$ and $e_6$ as pointing north (the edges look slightly more north than east). Edges $e_4$ and $e_5$, in turn, would be classified as pointing south (the edges look slightly more south than east).
The consistency of flow could then be calculated by dividing the number of edges that are consistent with the predominant flow direction (i.e., all edges pointing east), by the overall number of edges resulting into a consistency score of $\nicefrac{4}{8}=0.5$. The assignment of just one direction to an edge would result, against our intuition, in a relatively low consistency of flow. Assigning two directions to each edge, instead of one, allows to better reflect our intuition of consistency of flow. Edges $e_1$, $e_2$, $e_7$, and $e_8$ would still be classified as east. Edges $e_3$ and $e_6$, however, would be classified as both north-east and edges $e_4$ and $e_5$ would be classified as south-east. To calculate the consistency of flow we can now consider all edges that have one direction assigned that is consistent with the flow (i.e., all edges pointing towards north-east or south-east would be considered correct) and divide them by the overall number of edges. In this case, and in line with our intuition, we would obtain a consistency score of $\nicefrac{8}{8}=1$.

In the following, we describe more formally how metric $\cons(G)$ captures our intuition of how the consistency of flow should be calculated. The proposed metric assigns to each graph $G$ a value between 0 and 1 quantifying the degree of consistency.
For this, we assume that the graph $G$ has a predominant flow direction and we have to determine it. The set of all possible directions is $D = \{\north, \east, \south, \west \}$, whereby the selection of the most predominant flow direction is based on majority voting (i.e., the direction most edges belong to is considered as the predominant flow direction). Precondition for the majority voting is that we can identify the direction of each edge. The function $\dir : L_E \to \mathcal{P}(D)$\footnote{With $\mathcal{P}(S)$ we indicate the powerset (i.e., the set of all subsets) of $S$.}, given the layout information of an edge, returns the set of directions (i.e., potentially more than one) the edge belongs to.
In Sections~\ref{sec:maa} we present the na\"ive approach to compute $\dir$, only considering one direction (which will serve as a baseline). In Section~\ref{sec:mab}, in turn, we describe the classification of edges into two directions.
We can then calculate the overall consistency $\cons(G)$ by dividing the number of edges belonging to the predominant flow direction by the number of edges.

Algorithm~\ref{alg:metric} highlights the calculation of the metric. The procedure requires a graph $G$ as input (with the layout details) and a $\dir$ function, in order to get the directions of an edge. The procedure then iterates through each edge (line~\ref{alg:metric:edge-iterate}) and adds its directions to the proper direction counters (line~\ref{alg:metric:add}). Frequency of the predominant direction is computed (line~\ref{alg:metric:count-best}) and the final result is returned (line~\ref{alg:metric:result}).
\begin{algorithm2e}[t]
	\DontPrintSemicolon
	\KwIn{$G = (V, E, L_V, L_E)$: graph with the representation of the process; $\dir$: a function to obtain the direction(s) of an edge}
	\BlankLine
	\tcc{Define the directions, and initialize one counter for each direction}
	$\textit{freqs}[\north] \gets 0$ \;
	$\textit{freqs}[\east] \gets 0$ \;
	$\textit{freqs}[\south] \gets 0$ \;
	$\textit{freqs}[\west] \gets 0$ \;
	\BlankLine
	\tcc{Iterate through all edges to populate $\textit{freqs}$}
	\For{$e \in E$\label{alg:metric:edge-iterate}} {
		\tcc{Contribution of the edge to each direction}
		$\textit{dirs}_e \gets \dir(e)$ \tcc{$\textit{dirs}$ is a set with all directions the edge $e$ is pointing to}
		\For{$d \in \{\north, \east, \south, \west \}$} {
			\tcc{If the direction $d$ is one of edge's direction, then increment the corresponding counter}
			\If{$d \in \textit{dirs}_e$} {
				$\textit{freqs}[d] \gets \textit{freqs}[d] + 1$ \label{alg:metric:add} \tcc{The same edge is allowed to belong to more than one direction}
			}
		}
	}
	\BlankLine
	\tcc{Obtain the cost of the predominant direction }
	$\textit{predominant} \gets \max\left\{ \textit{freqs}[\north], \textit{freqs}[\east], \textit{freqs}[\south], \textit{freqs}[\west] \right\}$ \label{alg:metric:count-best} \;
	\BlankLine
	\tcc{Return the final consistency score, assuming the graph has at least one edge (and therefore $|E| > 0$}
	\Return $\nicefrac{\textit{predominant}}{|E|}$ \label{alg:metric:result} \;
	\caption{Algorithmic specification of $\cons(G)$. \label{alg:metric}}
\end{algorithm2e}

The two upcoming subsections explain in details the two possible instantiations of the $\dir$ function. The first, assigns to each edge exactly one direction; the second assigns two directions to each edge. Since these two definitions of $\dir$ affect the final result of the $\cons$ function, we assign to the two possible combination of $\cons$ and $\dir$ different names: \mAa{} and \mAb.

\subsubsection{One Direction per Edge (\mAa)}
\label{sec:maa}

The first instance of the $\dir$ function we analyze, which might be considered as a na\"ive approach, assigns one direction to each edge. For readability purposes, in the rest of this paper, we refer to the $\cons$ function using the $\dir$ described in this section as \mAa.

\begin{figure}[t]
	\centering
	\begin{subfigure}[b]{.45\textwidth}
		\centering
		\begin{tikzpicture}[scale=0.9, every node/.style={scale=0.6}]
			\draw[step=5mm, gray!50, style=densely dotted] (-3,-2.5) grid (3,2.5);
			\draw[pattern=crosshatch dots, pattern color=black] (0,0) -- (1.05,1.05) arc(45:-45:1.5);
			\draw[fill=gray!50] (0,0) -- (0.7,0.7) arc(45:135:1);
			\draw[pattern=crosshatch, pattern color=black] (0,0) -- (-0.7,-0.7) arc(-135:-45:1);
			\draw[pattern=north west lines, pattern color=black] (0,0) -- (-1.05,1.05) arc(135:225:1.5);
			\draw[black] (-2,0) -- (2,0);
			\draw[black] (0,-2)--(0,2);
			\draw[black] (-1.5,-1.5) -- (1.5,1.5);
			\draw[black] (-1.5,1.5) -- (1.5,-1.5);
			\filldraw[red] (2,0) circle (1pt) node [right, black] {$0\degree$};
			\filldraw[red] (1.5,1.5) circle (1pt) node [above right, black] {$-45\degree$};
			\filldraw[red] (0,2) circle (1pt) node [above, black] {$-90\degree$};
			\filldraw[red] (-1.5,1.5) circle (1pt) node [above left, black] {$-135\degree$};
			\filldraw[red] (-2,0) circle (1pt) node [above, black] {$-180\degree$};
			\filldraw[red] (-2,0) circle (1pt) node [below, black] {$180\degree$};
			\filldraw[red] (-1.5,-1.5) circle (1pt) node [below left, black] {$135\degree$};
			\filldraw[red] (0,-2) circle (1pt) node [below, black] {$90\degree$};
			\filldraw[red] (1.5,-1.5) circle (1pt) node [below right, black] {$45\degree$};	
		\end{tikzpicture}
		\caption{Division of the radius to assign one direction to each edge (\mAa).}
		\label{fig:angles:one}
	\end{subfigure}
	\hfill
	\begin{subfigure}[b]{.45\textwidth}
		\centering
		\begin{tikzpicture}[scale=0.9, every node/.style={scale=0.6}]
			\draw[step=5mm, gray!50, style=densely dotted] (-3,-2.5) grid (3,2.5);
			\draw[pattern=crosshatch dots, pattern color=black] (0,1.5) arc(90:-90:1.5);
			\draw[pattern=north west lines, pattern color=black] (0,1.5) arc(90:270:1.5);
			\draw[fill=gray!50] (-1,0) arc(180:0:1);
			\draw[pattern=crosshatch, pattern color=black] (0,0) -- (-1,0) arc(-180:0:1);
			\draw[black] (-2,0) -- (2,0);
			\draw[black] (0,-2)--(0,2);
			\draw[black] (-1.5,-1.5) -- (1.5,1.5);
			\draw[black] (-1.5,1.5) -- (1.5,-1.5);
			\filldraw[red] (2,0) circle (1pt) node [right, black] {$0\degree$};
			\filldraw[red] (1.5,1.5) circle (1pt) node [above right, black] {$-45\degree$};
			\filldraw[red] (0,2) circle (1pt) node [above, black] {$-90\degree$};
			\filldraw[red] (-1.5,1.5) circle (1pt) node [above left, black] {$-135\degree$};
			\filldraw[red] (-2,0) circle (1pt) node [above, black] {$-180\degree$};
			\filldraw[red] (-2,0) circle (1pt) node [below, black] {$180\degree$};
			\filldraw[red] (-1.5,-1.5) circle (1pt) node [below left, black] {$135\degree$};
			\filldraw[red] (0,-2) circle (1pt) node [below, black] {$90\degree$};
			\filldraw[red] (1.5,-1.5) circle (1pt) node [below right, black] {$45\degree$};	
		\end{tikzpicture}
		\caption{Division of the radius to assign two directions to each edge (\mAb).}
		\label{fig:angles:two}
	\end{subfigure}
	\caption{Division of the radius according to the number of directions per edge that could be defined. These two cases report $\north$ (gray filled area), $\east$ (dotted area), $\south$ (grid area) and $\west$ (lined area) directions.}
\end{figure}
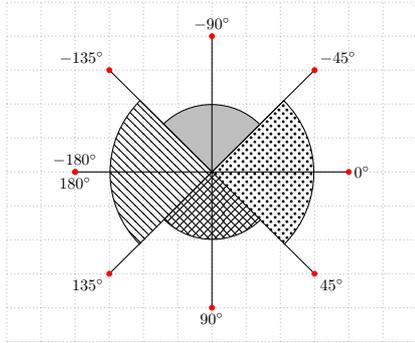
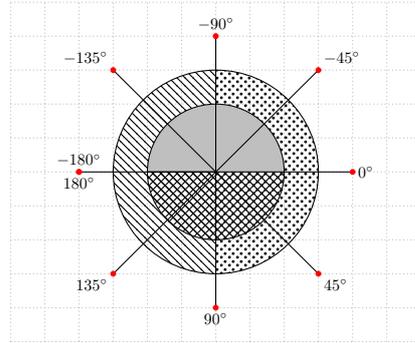

In order to identify the directions of each edge, we consider the ``angle'' created by the edge.
Starting from the coordinates of the start and end points of an edge and using the arctangent function with two arguments, we can get the actual angle of the edge. To determine the direction of the edge we divide the radius into four equal parts of $90\degree$ (one for each direction, i.e., $\north, \east, \south, \west$).  Figure~\ref{fig:angles:one} highlights the four directions: the filled area identifies the $\north$ direction; the dotted area identifies $\east$, the grid area represents $\south$, and the lined area identifies the $\west$ direction.

We then check whether the angle obtained for a particular edge is within the intervals referring to each direction. Since the angles corresponding to each direction do not overlap, $\dir$ always assigns exactly one direction to each edge.

\subsubsection*{Examples}

In the following we illustrate the calculation of the M-E1 metric using the examples shown in Fig.~\ref{fig:models-examples}. Table~\ref{tbl:scores:one-direction} summarizes the obtained results.
For example, if we consider the process of Figure~\ref{fig:models-examples:linear}, we can see that it contains
1 edges looking $\north$,
48 looking $\east$,
2 looking $\west$, and
0 looking $\south$.
As reported in line~\ref{alg:metric:result} of Alg.~\ref{alg:metric}, the final score is given by $\nicefrac{\textit{predominant}}{|E|}$, where $\textit{predominant}$ is the frequency of the prevalent direction ($\east$ in our case, with frequency 48) and $|E|$ is the number of edges of the graph.
Therefore, if we apply these counters, the final consistency score for this model is $\nicefrac{48}{51}=0.941$.

In Figure~\ref{fig:models-examples:fragment} we have
1 edge looking $\north$,
50 looking $\east$,
4 looking $\west$, and
4 looking $\south$. Therefore, the final consistency score is $\nicefrac{50}{59}=0.847$.

Finally, in Figure~\ref{fig:models-examples:messy} we have
5 edges looking $\north$,
20 looking $\east$,
17 looking $\west$, and
9 looking $\south$. Therefore, the final consistency score is $\nicefrac{20}{51}=0.392$.
\begin{table}
	\centering
	\begin{tabular}{lr}
		\toprule
		\textbf{Model} & \textbf{Consistency using \mAa} \\
		\midrule
		Figure~\ref{fig:models-examples:linear}         & 0.941 \\
		Figure~\ref{fig:models-examples:fragment}       & 0.847 \\
		Figure~\ref{fig:models-examples:messy}          & 0.392 \\
		\bottomrule
	\end{tabular}
	\caption{Consistency scores computed using the \mAa{} metric.}
	\label{tbl:scores:one-direction}
\end{table}

\subsubsection{Two Directions per Edge (\mAb)}
\label{sec:mab}

In this subsection we are going to report details regarding the second definition of the $\dir$ function. This new version of the function assigns two directions to each edge. For readability purposes, in the rest of this paper, we refer to the $\cons$ function using the $\dir$ described in this section as \mAb.

The main difference, with respect to the definition provided before is that now each direction corresponds to a $180\degree$ portion of the angle. Based on this definition of direction, each portion overlaps with two others (cf. Figure~\ref{fig:angles:two}). In this case, it is possible to see that the $\east$ direction (dotted area) overlaps with both the $\north$ (filled area) and the $\south$ (grid area) directions. The result is that, with this metric, each edge is always associated to exactly two directions.

\subsubsection*{Examples}

\begin{table}
	\centering
	\begin{tabular}{lr}
		\toprule
		\textbf{Model} & \textbf{Consistency using \mAb} \\
		\midrule
		Figure~\ref{fig:models-examples:linear}         & 0.960 \\
		Figure~\ref{fig:models-examples:fragment}       & 0.915 \\
		Figure~\ref{fig:models-examples:messy}          & 0.588 \\
		\bottomrule
	\end{tabular}
	\caption{Consistency scores computed using the \mAb{} metric.}
	\label{tbl:scores:two-directions}
\end{table}

Despite this slight change, the overall metric could generate very different values. For example, if we consider again the process models seen so far, we can observe the score values reported in Table~\ref{tbl:scores:two-directions}.

The process of Figure~\ref{fig:models-examples:linear} contains
28 edges looking $\north$,
23 pointing $\south$,
49 pointing $\east$, and
2 looking $\west$,
with the predominant flow direction $\east$.
Therefore, if we compute the same values of Alg.~\ref{alg:metric}, the final consistency score for this model is $\nicefrac{49}{51}=0.960$.

In Figure~\ref{fig:models-examples:fragment} we have
28 edges looking $\north$,
31 pointing $\south$,
54 pointing $\east$, and
5 looking $\west$,
again with the predominant flow direction $\east$.
Therefore, the final consistency score is $\nicefrac{54}{59}=0.915$.

Finally, in Figure~\ref{fig:models-examples:messy} we have
21 edges looking $\north$,
30 pointing $\south$,
27 pointing $\east$, and
24 looking $\west$.
Therefore, the final consistency score is $\nicefrac{30}{51}=0.588$.
Unlike with previous examples for this model the predominant flow direction is $\south$.

\vspace{1em}
Please note that the time complexity of both the $\dir$ procedures reported in the last two subsections is constant, given an edge. Therefore, the general complexity of the $\cons$ function is linear on the number of edges of the graph.
Another key characteristic of these metrics is their semantic independence, i.e., they can be applied to any directed graph.

\subsection{Metric Based on Behavioral Profiles (\mB)}

While the two metrics introduced in Section~\ref{sec:edges-direction-metrics} consider the edges of a process model, this metric puts its focus on activities to determine the extent to which the layout of the model is consistent with the temporal logical ordering of the business process. For this, the metric looks at the relations between pairs of activities (i.e., their \emph{behavioral profiles}~\cite{Weidlich2011}) and evaluates, for each of them,
whether the way they are placed in relation to each other is consistent with their temporal logical ordering. For readability purposes, in the rest of this paper, we refer to the metric described in this section as \mB.

\begin{figure}
	\includegraphics[width=\textwidth]{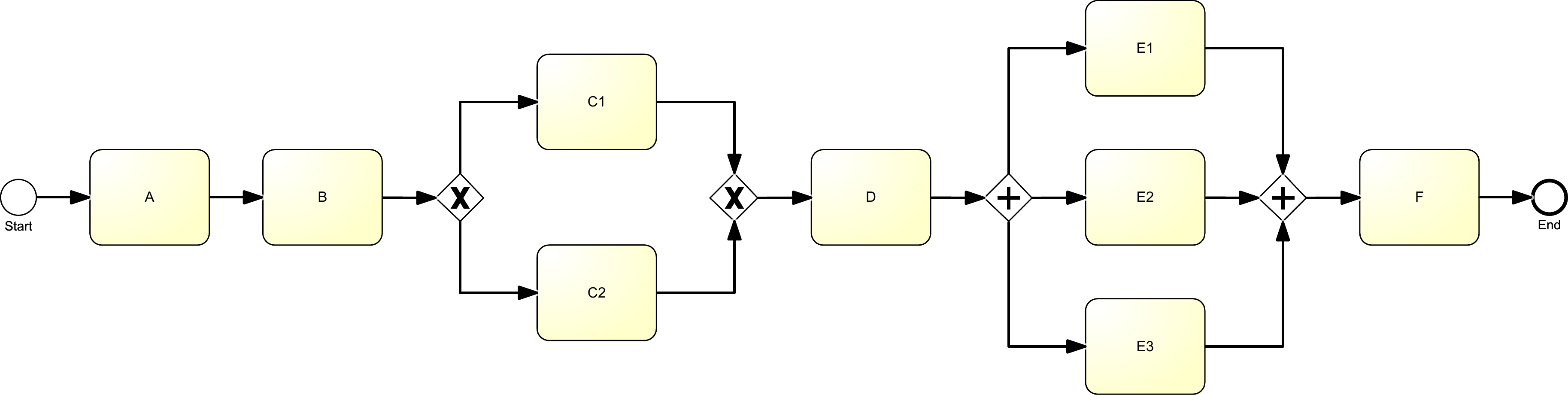}
	\caption{Example of process model for illustrating behavioral profiles between activities.}
	\label{fig:bpmn-bp}
\end{figure}

The fundamental idea behind behavioral profiles consists of the characterization of a process using relations between two activities, defined according to three fundamental possibilities:
\emph{(i)} strict order;
\emph{(ii)} exclusiveness; or
\emph{(iii)} interleaving order.
Let us present these relations using the example process model depicted in Figure~\ref{fig:bpmn-bp}. Between activities $A$ and $B$ the \emph{strict order} is holding, identified as $A \to B$ (i.e., $A$ always occurs before $B$, and never the other way round). Activities $C1$ and $C2$ are in as \emph{exclusiveness} relation $C1 + C2$ (i.e., $C1$ cannot appear before $C2$ and $C2$ cannot appear before $C1$). Finally, $E1$ and $E2$ (but also $E3$) are in \emph{interleaving order}: $E1 \| E2$ (i.e., $E1$ might appear before $E2$ and $E2$ might appear before $E1$ as well).

The main idea of the metric \mB{} is to measure the extent to which the layout of a process model reflects the temporal logical ordering of the activities.
The behavioral relation that imposes a restrictive order among activities is the \emph{strict order behavioral relation}. Therefore, we need to analyze the position of nodes referring to activities belonging to such relation. Then, for each strict order relation, we check whether the source node (i.e., the node that must occur first) is ``graphically before'' the target node (i.e., the node that must occur later).
The ``graphically before'' condition holds if the target node is placed east or south of the source node, i.e., the positioning of the two activities reflects their temporal logical ordering. To calculate the consistency score we divide the number of graphically before relations by the overall number of strict relations.

More formally, given a process graph $G = (V, E, L_V, L_E)$, let us define a behavioral relationship $b$ as the tuple $b=\langle r, s, t \rangle$ with $r \in \{ \to, +, \| \}$ indicating the relation type and $s,t \in V$ indicating the nodes associated to the source and the target activities (therefore, $s$ and $t$ must refer to nodes representing activities, not just ``general nodes'' of $G$).
For convenience, we define projection operators for a behavioral relation instance $b=\langle r, s, t \rangle$ such that $\#_\textit{relation}(b) = r$, $\#_\textit{source}(b) = s$, and $\#_\textit{target}(b) = t$.
We also assume to have available a procedure $\bp$ which extracts all behavioral relations out of a process.\footnote{In this work we assume to compute the behavioral profiles using \emph{look ahead} value 1.} The complete pseudocode of the algorithm is reported in Algorithm~\ref{alg:direction-bp}.
\begin{algorithm2e}[t]
	\DontPrintSemicolon
	\KwIn{$G = (V, E, L_V, L_E)$: }
	\BlankLine
	$t_{\small\textit{strict}} \gets 0$ \;
	$\textit{correct}_{\small\east} \gets 0$ \;
	$\textit{correct}_{\small\south} \gets 0$ \;
	$\textit{BP} \gets \bp(G)$ \tcc{Compute all behavioral relations} \label{alg:direction-bp:bp}
	\ForEach{$\textit{bp} \in \textit{BP}$} {
		\If(\tcc*[h]{Only strict order relations}\label{alg:direction-bp:filter}){$\#_\textit{relation}(\textit{bp}) = \to$} {
			\BlankLine
			\tcc{Extract the coordinates of the central points of the source and target nodes}
			$(s_x, s_y) \gets L_V(\#_\textit{source}(\textit{bp}))$ \; \label{alg:direction-bp:source}
			$(t_x, t_y) \gets L_V(\#_\textit{target}(\textit{bp}))$ \;\label{alg:direction-bp:target}
			\BlankLine
			\If(\tcc*[h]{Check for the $\east$ direction}\label{alg:direction-bp:startcons}){$s_x < t_x$} {
				$\textit{correct}_{\small\east} \gets \textit{correct}_{\small\east} + 1$ \;
			}
			\If(\tcc*[h]{Check for the $\south$ direction}){$s_y < t_y$} {
				$\textit{correct}_{\small\south} \gets \textit{correct}_{\small\south} + 1$ \;
			}\label{alg:direction-bp:endcons}
			\BlankLine
			$t_{\small\textit{strict}} \gets t_{\small\textit{strict}} + 1$ \;
		}
	}
	\Return $\max\left\{ \textit{correct}_{\small\east}, \textit{correct}_{\small\south} \right\} / t_{\small\textit{strict}}$ \tcc{Final consistency score as the dominant direction, divided by the total number of strict relations} \label{alg:direction-bp:final}
	\caption{Specification of metric for the computation of the layout consistency using  the behvioral relations. \label{alg:direction-bp}}
\end{algorithm2e}
The algorithm starts by initializing several counters, and by extracting all the behavioral profiles available in the process (line~\ref{alg:direction-bp:bp}). Then, it has to iterate through all relations and consider only the strict orders (line~\ref{alg:direction-bp:filter}). For these relations the coordinates of source and start activities are extracted (lines~\ref{alg:direction-bp:source}-\ref{alg:direction-bp:target}) and the system checks whether the graphically before condition holds (lines~\ref{alg:direction-bp:startcons}-\ref{alg:direction-bp:endcons}). The final score is computed as the ratio of the graphically before relations divided by the total number of strict relations (line~\ref{alg:direction-bp:final}).

\vspace{1em}
The complexity of the algorithm is linear on the number of behavioral relations that could be extracted. Behavioral profiles can be computed quite efficiently, in a low polynomial time to the size of the process model \cite{We2011}.

\subsubsection*{Examples}

\begin{table}
	\centering
	\begin{tabular}{lr}
		\toprule
		\textbf{Model} & \textbf{Consistency using \mB} \\
		\midrule
		Figure~\ref{fig:models-examples:linear}         & 0.930 \\
		Figure~\ref{fig:models-examples:fragment}       & 0.868 \\
		Figure~\ref{fig:models-examples:messy}          & 0.622 \\
		\bottomrule
	\end{tabular}
	\caption{Consistency scores computed using the \mB{} metric.}
	\label{tbl:scores:bp}
\end{table}

If we consider the example process models seen so far, we can observe the score values reported in Table~\ref{tbl:scores:bp}.
For example, the process of Figure~\ref{fig:models-examples:linear} has
43 strict relations (computed with \emph{look ahead} 1),
and 40 of them are pointing $\south-\east$.
Therefore, the final score is $\nicefrac{40}{43}=0.930$.
In Figure~\ref{fig:models-examples:fragment} we have
38 strict relations (with \emph{look ahead} 1),
and 33 of them are pointing $\south-\east$.
Therefore, the final score is $\nicefrac{33}{38}=0.868$.
Finally, in Figure~\ref{fig:models-examples:messy} we have
37 strict relations (with \emph{look ahead} 1),
and 23 of them are pointing $\south-\east$.
Therefore, the final score is $\nicefrac{23}{37}=0.622$.

Please note that, since we are using \emph{look ahead} equals to 1, we do not penalize the violation of the ``graphically before'' condition for relations involving activities very far apart in the process. In particular, value 1 for the \emph{look ahead} indicates that only pairs of very close activities are considered.
Although this is a parameter of our approach (i.e., it can be changed), we opted for this configuration since each local violation implies a change in direction and, this way, we count the changes without penalizing more than once for each change.

\section{Evaluation of Flow Consistency}
\label{sec:evaluation}

To gain a better understanding of our new flow consistency metrics, we conducted several empirical evaluations. In total, we performed 3 evaluations, useful to provide a more comprehensive picture of the capabilities of our metrics.

Since the approaches we defined rely on different assumptions, and therefore quantify the flow consistency using different feature sets, with the first evaluation we wanted to establish to what extent the metrics ``agree'' on the same model.
We can use this analysis to ensure that the features we are taking into account are not redundant to each other and therefore be sure that our approaches are measuring the consistency of the flow considering different perspectives.
If this is the case, we could identify the conditions under which a metric performs better than another one.
Another evaluation focused only on the time efficiency of our metrics, i.e., the time required to compute each metric on all the models of our dataset.
The third evaluation is an experimental one, performed with the support of several people familiar with process modeling, and aims at comparing the human assessment of flow consistency with the outcomes provided by our metrics.

Both automated analyses are based on a process model dataset which was generated during a modeling session conducted in December 2012 at the Eindhoven University of Technology, with students following programs on operations management and logistics, business information systems, innovation management, and human-technology interaction \cite{Pinggera2014}. In total, the dataset contains 125 models, all referring to the same process description. The experimental evaluation is based on a subset of this dataset containing 14 models.

\subsection{Metrics Agreement}
\label{sec:metrics-analysis}

The first analysis we performed aimed at establishing the extent to which different metrics agree. In order to evaluate this aspect, we counted the number of models that each metric places within a provided consistency score interval.
\begin{figure}
	\includegraphics[trim={2.5cm 5cm 2.5cm 5cm},width=\textwidth]{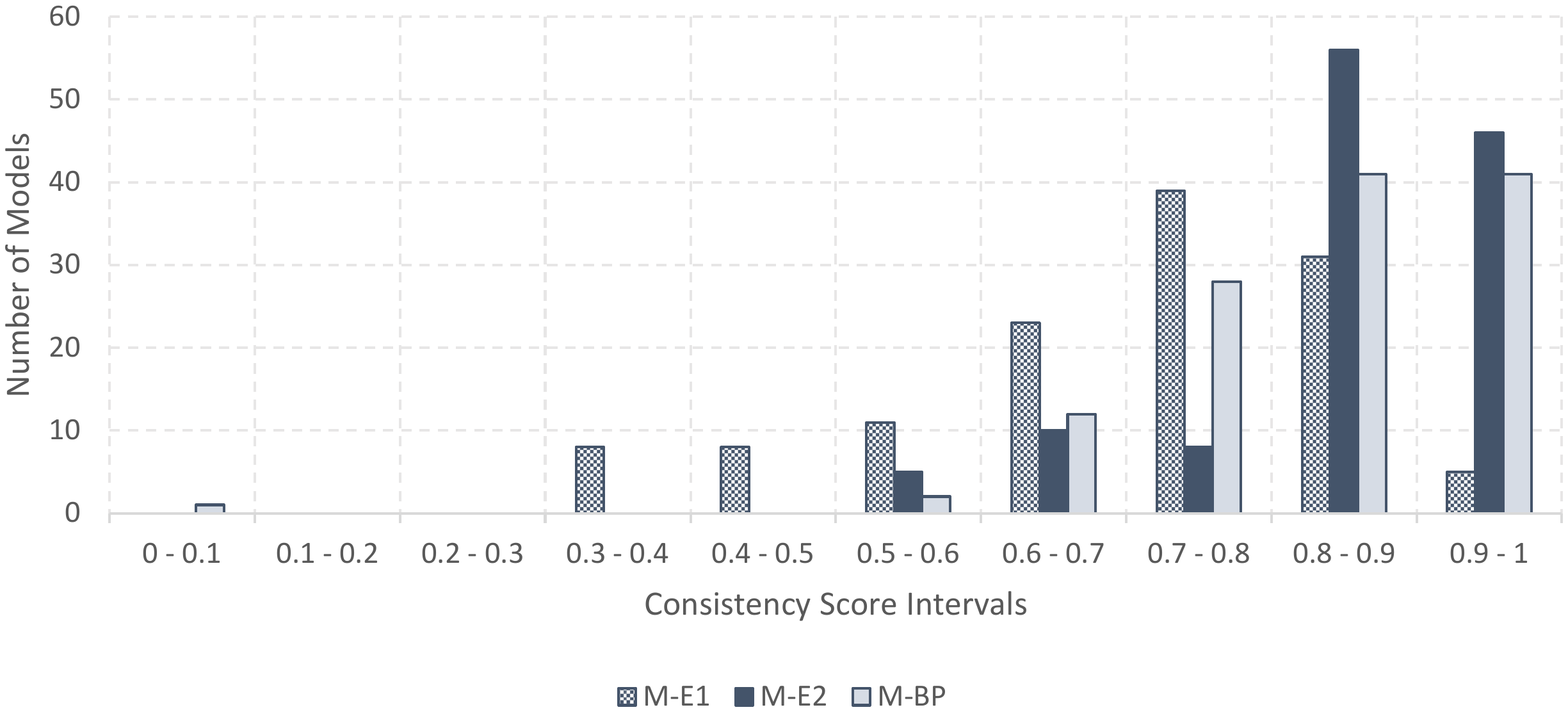}
	\caption{Number of process models within different consistency score intervals.}
	\label{fig:score-distribution}
\end{figure}
Figure~\ref{fig:score-distribution} depicts the distribution of the values, using intervals of 0.1. It is interesting to note that the metrics tend to assign scores in slightly different intervals. For example, \mAa{} distributes the scores rather uniformly, but there is a considerable set of models (above the average) with scores in the interval $0.6-0.9$. Metric \mB{}, in turn, assigns rather high values, with only very few exceptions below $0.6$. Finally, metric \mAb{} assigns even higher scores, and most models lie in the interval $0.8-1$.

In order to compare the agreement of our metrics, we decided to use ranking. Specifically, using the scores generated by each metric we ranked the dataset, from the most consistent model to the least consistent one. Therefore, for each model, we ended up with three ranking positions (one for each metric). We computed the average and the standard deviation of the rankings for each model and we plot the latter value.
%
\begin{figure}
	\includegraphics[trim={2.5cm 5cm 2.5cm 5cm},width=\textwidth]{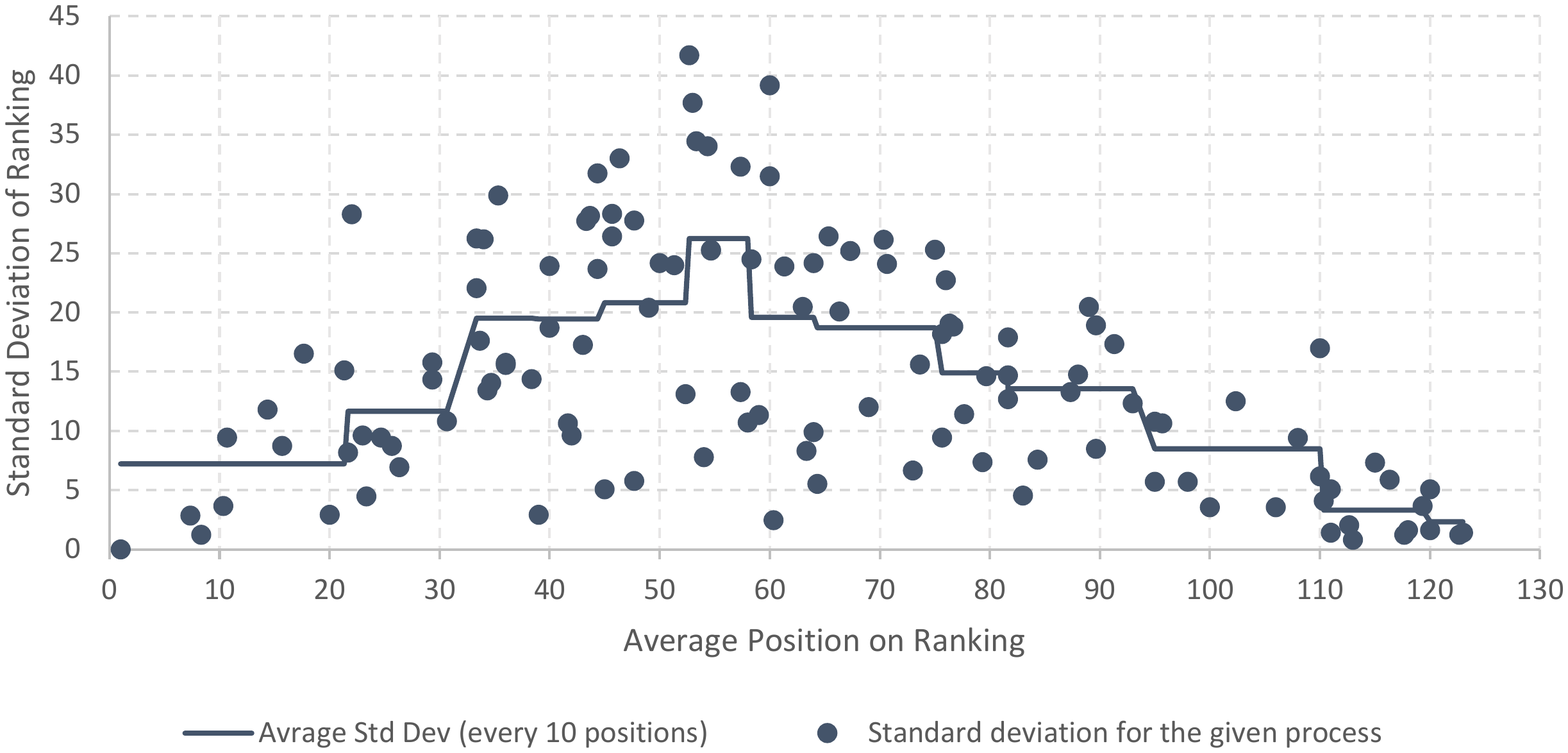}
	\caption{
		In this chart each dot represents a process model. The position on the $x$ axis indicates its average ranking; the position on the $y$ axis indicates the standard deviation of the rankings. Both averages and standard deviations are computed for all three metrics.}
	\label{fig:stddev-distribution}
\end{figure}
Figure~\ref{fig:stddev-distribution} contains the results of our analysis. The figure does not only show the evolution of the standard deviation as the average ranking increases, but it also reports the average values, computed every 10 data points.
By looking at the average values of the standard deviation, it is interesting to highlight that our metrics tend to agree with the ranking next to the extremes of the chart (i.e., lower standard deviation averages for very high or very low rankings). Instead, on the middle ranking positions, values are more spread apart: the peak of the average standard deviation is reached for ranking positions $52-58$.
This clearly indicates that all metrics, basically, agree on ``very consistent'' and ``very inconsistent'' models. However, for average process quality scenarios, there is less agreement.

The observed behavior is in line with our expectations since very consistent and very inconsistent models are ``globally'' recognized, no matter what the considered features are. This is highlighted by our observation: our metrics (and the features taken into account) capture very consistent or very inconsistent scenarios similarly (i.e., low standard deviations on ranking).
Moreover, for average situations (i.e., average rankings), the characteristics of each metric play an important role: the different feature sets used indeed provide different characterizations which, in turn, results in values more spread apart.
These large standard deviations, on average cases, also indicate that the specific metric's features are not redundant to each other.

\vspace{1em}
Consider as example the linear process model illustrated in Fig.~\ref{fig:models-examples:linear}. It is ranked in position 3, 5 and 24 (respectively by \mAa{}, \mAb{}, and \mB{} metrics), with a rather low standard deviation of 9.4. This is an example of a model that all metrics consider as a consistent one. It is placed on the very left side of the chart in Fig.~\ref{fig:stddev-distribution}.

The model reported in Fig.~\ref{fig:models-examples:messy} has an even lower standard deviation (1.63) with the following rankings: 118 for \mAa{}, 122 for \mAb{}, and 120 for \mB{}. This indicates that all metrics consider this an inconsistent model. It is located on the very right side of Fig.~\ref{fig:stddev-distribution}.

The model depicted in Fig.~\ref{fig:models-examples:fragment} is not consistently ranked by all metrics (positions 11 for \mAa{}, 36 for \mAb{}, and 54 for \mB{}). This model has a standard deviation of 17.6 and appears towards the central part of Fig.~\ref{fig:stddev-distribution}. This model clearly highlights the features that each metric takes into account: edges' direction and behavioral profiles violations. In particular, considering the directions of edges, this model, compared to other models, has only few inconsistencies (such as those connecting each horizontal fragment) and many properly oriented edges.
From a behavioral profiles perspective, instead, the model has a considerable amount of strict relations that are violating the ``graphically before'' condition (again, compared to the other models).

\vspace{1em}
To sum up, the metrics we described in this paper are providing consistent results regarding ``extremes models'' (i.e., models with very high and very low consistency scores). Instead, on average scenarios, it depends on what the end user wants to analyze: metric \mB{} is concentrating on the position of activities; \mAa{} and \mAb{} are based on the direction of edges.
In Section~\ref{sec:experimental-evaluation} we will investigate which representation is more in line with the human perception.

\subsection{Efficiency}

All our metrics have been implemented as extension of the Cheetah Experimental Platform \cite{Pinggera2010}. In order to evaluate the efficiency of these metrics, we compared the time required to compute them for each model. The machine we used is a standard-level laptop with a Intel Core i7-2620M CPU (2.70 GHz), equipped with 8GB of RAM and a 64bit Windows~7 OS. The test was performed on the Java Virtual Machine version 1.8.0.25, with 64bit.

\begin{table}
	\centering
	\begin{tabular}{r|r|r|r}
		\toprule
		& \textbf{\mAa{}} & \textbf{\mAb{}} & \textbf{\mB{}} \\
		\midrule
		\textbf{Average} (ms)	& 0.1533	& 0.0693	&  34.4179	\\
		\textbf{Max} (ms) &	2.0011	& 0.8164	& 174.4437 \\
		\textbf{Min} (ms)	& 0.0524	& 0.0161	&   2.4495 \\
		\bottomrule
	\end{tabular}
	\caption{Time required to compute the metrics for one process model.}
	\label{tbl:times}
\end{table}
Final results are reported in Table~\ref{tbl:times} (all values are expressed in milliseconds).
The results report the average, the minimum and the maximum computation time required to calculate the metrics for the whole dataset. To provide more general results and to avoid that specific conditions influence the computation, we computed each metric 5 times for each process (i.e., $5 \times 125 = 625$ computations for each metric) and the average times are reported.
As the complexity analyses suggested, all three metrics can be considered as very efficient. From all three metrics, the behavioral profiles-based metric, \mB{}, is the most demanding one, since it has to compute the behavioral profiles in advance. Still the metric can be calculated, on average, within 34 milliseconds.
These results clearly show the applicability of the proposed metrics even for settings with high performance requirements (e.g., online and real-time computations needed, for example, to include such calculations in an intelligent modeling environment that provides recommendations to users based on observed behavior).

\subsection{Experimental Evaluation}
\label{sec:experimental-evaluation}


The evaluation reported in Section~\ref{sec:metrics-analysis} showed that the three proposed metrics tend to agree on the assessment of process models with very high and very low consistency of flow. For models with average ratings the assessment is less consistent. In this section we aim to evaluate how far the proposed metrics reflect human perception of consistency of flow. To do that, we conducted an empirical study in which we asked human readers to rate a set of models regarding their flow consistency.

\paragraph{Subjects}

For our evaluations, we targeted subjects familiar with process modeling, most of them coming from academia. In particular, our subjects were participants attending BPM 2015 conference\footnote{See \url{http://bpm2015.q-e.at/} for more information.}, since we expected them to be familiar with process modeling. In total, we collected data from 47 subjects.

\paragraph{Objects}

\begin{figure}
	\includegraphics[trim={2.5cm 5cm 2.5cm 5cm},width=\textwidth]{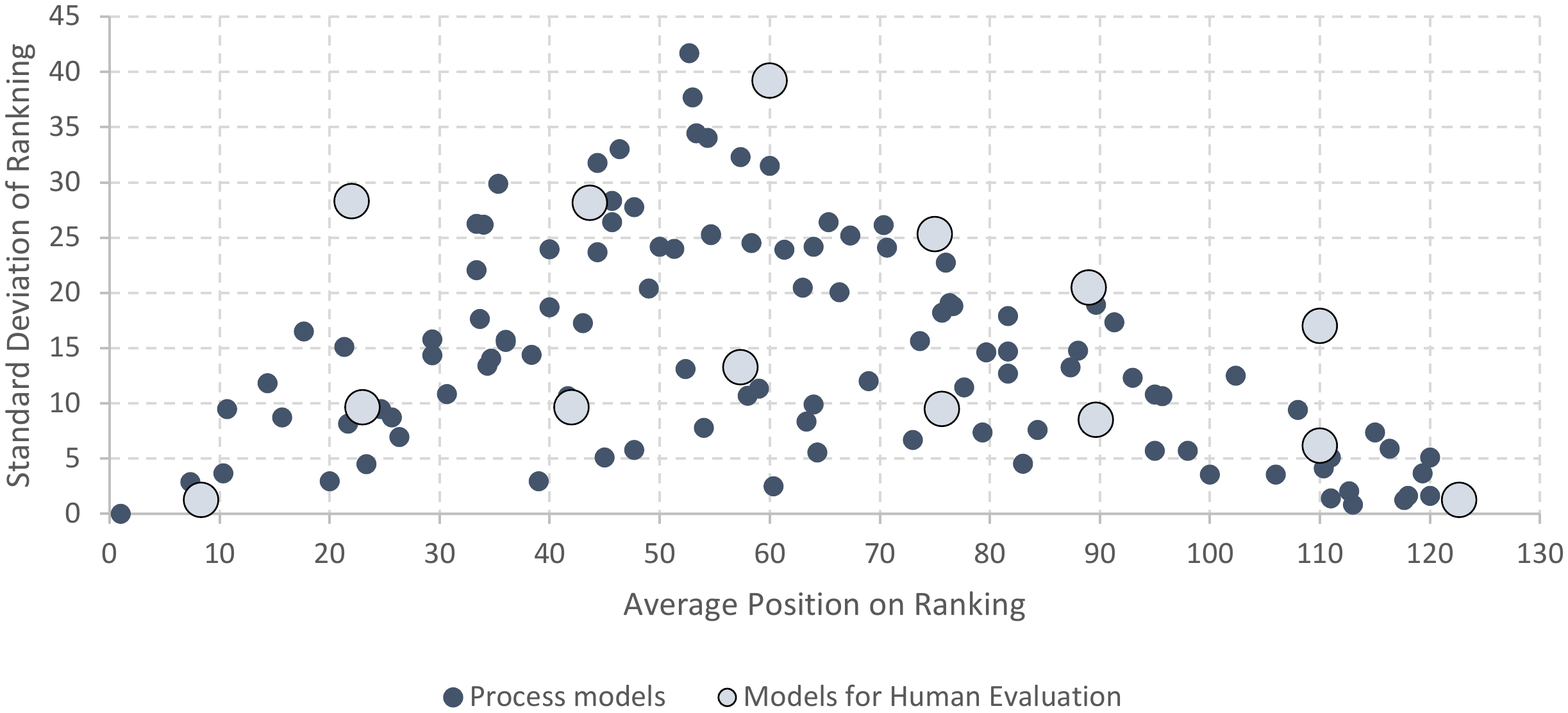}
	\caption{Average position on ranking with processes selected for the human assessment represented by large, bright-blue dots.}
	\label{fig:stddev-distribution-selected}
\end{figure}

We decided to select a subset of 14 models from our dataset. The models we picked have been sampled according to the standard deviation on the average ranking. In particular, we used the representation provided in Fig.~\ref{fig:stddev-distribution-selected}: we evenly sampled our space and, for each average ranking positions, two models are selected: one with low standard deviation and one with high standard deviation.
By doing that, we evenly partitioned our dataset, both with respect to the average ranking and the standard deviation. By selecting one process per partition, we guarantee that each partition is actually represented in our dataset.

We prepared a questionnaire for our subjects with the 14 models that were selected.\footnote{All the material used for this study is available at \url{http://bpm.q-e.at/experiment_flow_consistency}.}
We actually assembled two versions of the same questionnaire (labeled ``A'' and ``B'), with the processes presented in inverted order, to avoid that the ordering influences the evaluation.

\paragraph{Response Variables and Data Collection}

For each model reported in the questionnaire, we asked participants to rate the consistency of flow on a 7-point Likert scale ranging from 1 ``No consistency at all'' to 7 ``Complete consistency''.

\paragraph{Execution}

The actual evaluation was conducted in between August 31\textsuperscript{st} and September 3\textsuperscript{rd} 2015, during the BPM conference, which took place in Innsbruck. We asked all the conference participants to fill our questionnaire and return the answers to us, without providing them any additional instruction. We randomly distributed to each participant either a questionnaire labeled ``A'' or ``B''. In total we collected 47 answers (25 labeled ``A'' and 22 ``B'').


\paragraph{Data Analysis and Results}

Once we collected all our questionnaires we loaded the data into a spreadsheet. Then, we rescaled the values from the closed interval $[1,7]$ into $[0,1]$, just to simplify the comparison with the output of our metrics. For each model, we computed the average scores assigned by our participants. Then, we compared the average human evaluation against the automatic metrics defined in this paper.
The average human evaluation scores of all the models as well as the values of the three metrics are reported in Table~\ref{tbl:detailed-scores}.
%
\begin{table}
	\centering
	\begin{tabular}{r|rrr|r|r}
		\toprule
		& \multirow{2}{*}{\textbf{\mAa}} & \multirow{2}{*}{\textbf{\mAb}} &\multirow{2}{*}{\textbf{\mB}} & \multicolumn{2}{c}{\textbf{Human Evaluation}} \\
		& & & & \textbf{Average} & \textbf{St. Dev.} \\
		\midrule
		 Model 1 & 0.73 & 0.85 & 0.68 & 0.43 & 0.25 \\
		 Model 2 & 0.38 & 0.57 & 0.57 & 0.36 & 0.27 \\
		 Model 3 & 0.73 & 0.84 & 0.83 & 0.52 & 0.25 \\
		 Model 4 & 0.79 & 0.87 & 0.85 & 0.48 & 0.28 \\
		 Model 5 & 0.37 & 0.59 & 0.78 & 0.39 & 0.26 \\
		 Model 6 & 0.75 & 0.91 & 0.92 & 0.32 & 0.24 \\
		 Model 7 & 0.50 & 0.88 & 0.95 & 0.76 & 0.19 \\
		 Model 8 & 0.69 & 0.94 & 0.91 & 0.72 & 0.25 \\
		 Model 9 & 0.55 & 0.64 & 0.70 & 0.50 & 0.30 \\
		Model 10 & 0.86 & 0.92 & 0.93 & 0.73 & 0.20 \\
		Model 11 & 0.78 & 0.86 & 0.71 & 0.35 & 0.26 \\
		Model 12 & 0.74 & 0.96 & 1.00 & 0.80 & 0.19 \\
		Model 13 & 0.63 & 0.81 & 0.81 & 0.55 & 0.29 \\
		Model 14 & 0.87 & 0.96 & 0.97 & 0.66 & 0.25 \\
		\bottomrule
	\end{tabular}
	\caption{Descriptive variables for every model of the study. Scores for the three metrics are reported, together with the average score, and the standard deviation of the human assessment.}
	\label{tbl:detailed-scores}
\end{table}

The first test we employed was the One-Sample Kolmogorov-Smirnov, in order to verify the normality of our data distribution which is precondition to compute the Pearson correlation. The data we collected, actually, are fitting a normal distribution with mean 0.541 and standard deviation 0.16. The significance score observed is 0.919.

\begin{table}
	\centering
	\begin{tabular}{l|ccc}
		\toprule
		& \textbf{\mAa} & \textbf{\mAb} & \textbf{\mB} \\
		\midrule
		\textbf{Pearson Correlation}  & 0.263 & 0.567 & \textbf{0.719} \\
		\textbf{Significance} & 0.364 & 0.034 & \textbf{0.004} \\
		\midrule
	\end{tabular}
	\caption{Correlation of our three different approaches computed with average scores assigned by our subjects.}
	\label{tbl:correlation}
\end{table}
Once we verified such condition, we computed the Pearson correlation between each metric and the average human scores. The results are reported in Table~\ref{tbl:correlation}.
Our results suggest that the metric based on behavioral profiles, \mB{}, best reflects human assessment, with a Pearson correlation score of 0.719 and a significance value of 0.004. Please note that the absolute scores of the human evaluation and \mB{} are linearly shifted by a factor which, on average, is 0.29.
Also metric \mAb{} obtained a significant correlation with the human judgment (Pearson correlation 0.567, significance 0.034), but when compared to metric \mB{} the correlation is less strong.
Metric \mAa{}, in turn, with a Pearson correlation of 0.263 (and significance 0.364) is not related to human judgment.
\begin{figure}[t]
	\centering
	\begin{subfigure}[b]{.49\textwidth}
		\includegraphics[trim={5cm 2cm 3cm 2cm},width=\textwidth]{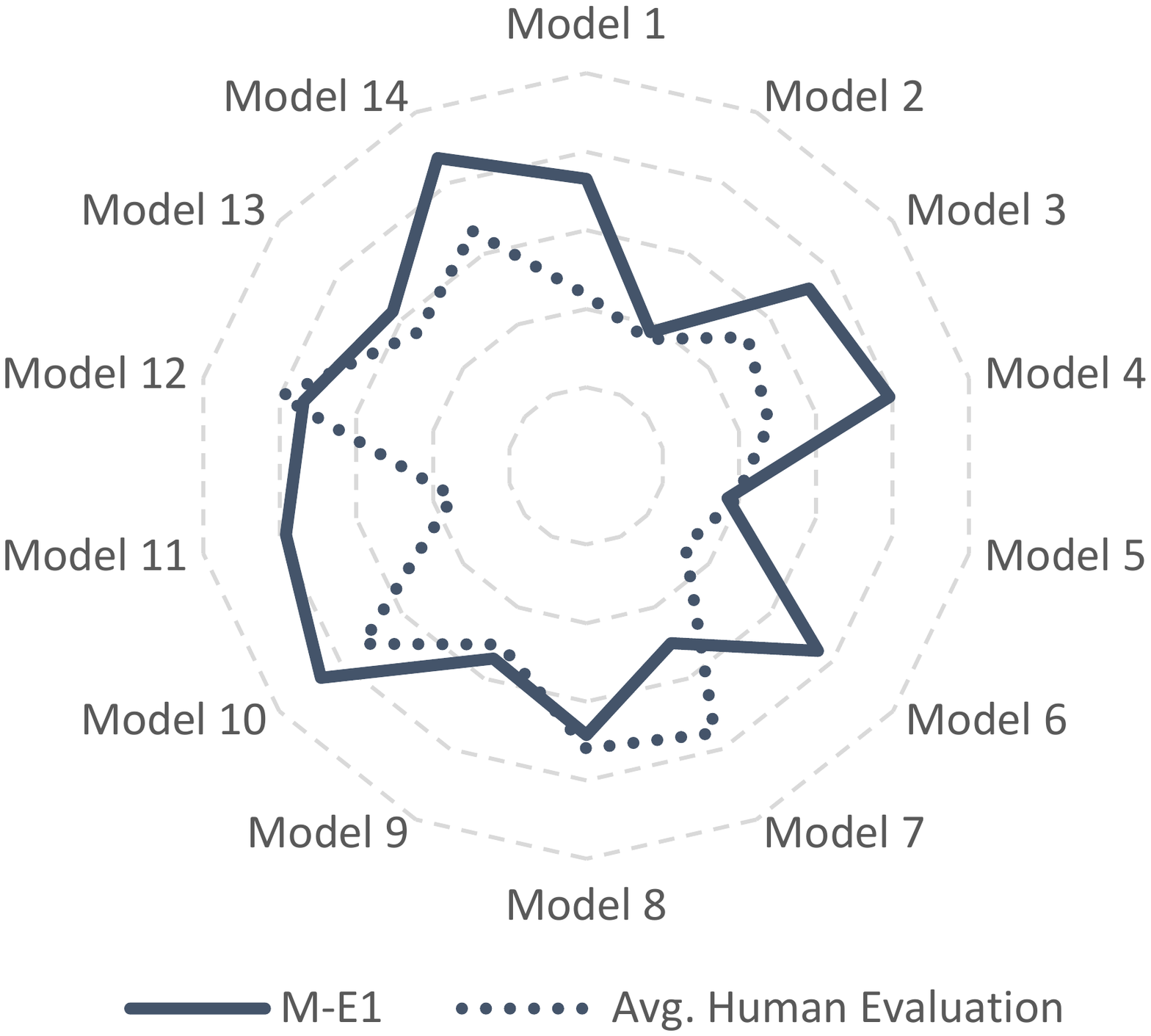}
		\caption{\mAa{} vs. human evaluation. \label{fig:score-trends:maa}}
	\end{subfigure}
	\begin{subfigure}[b]{.49\textwidth}
		\includegraphics[trim={3cm 2cm 5cm 2cm},width=\textwidth]{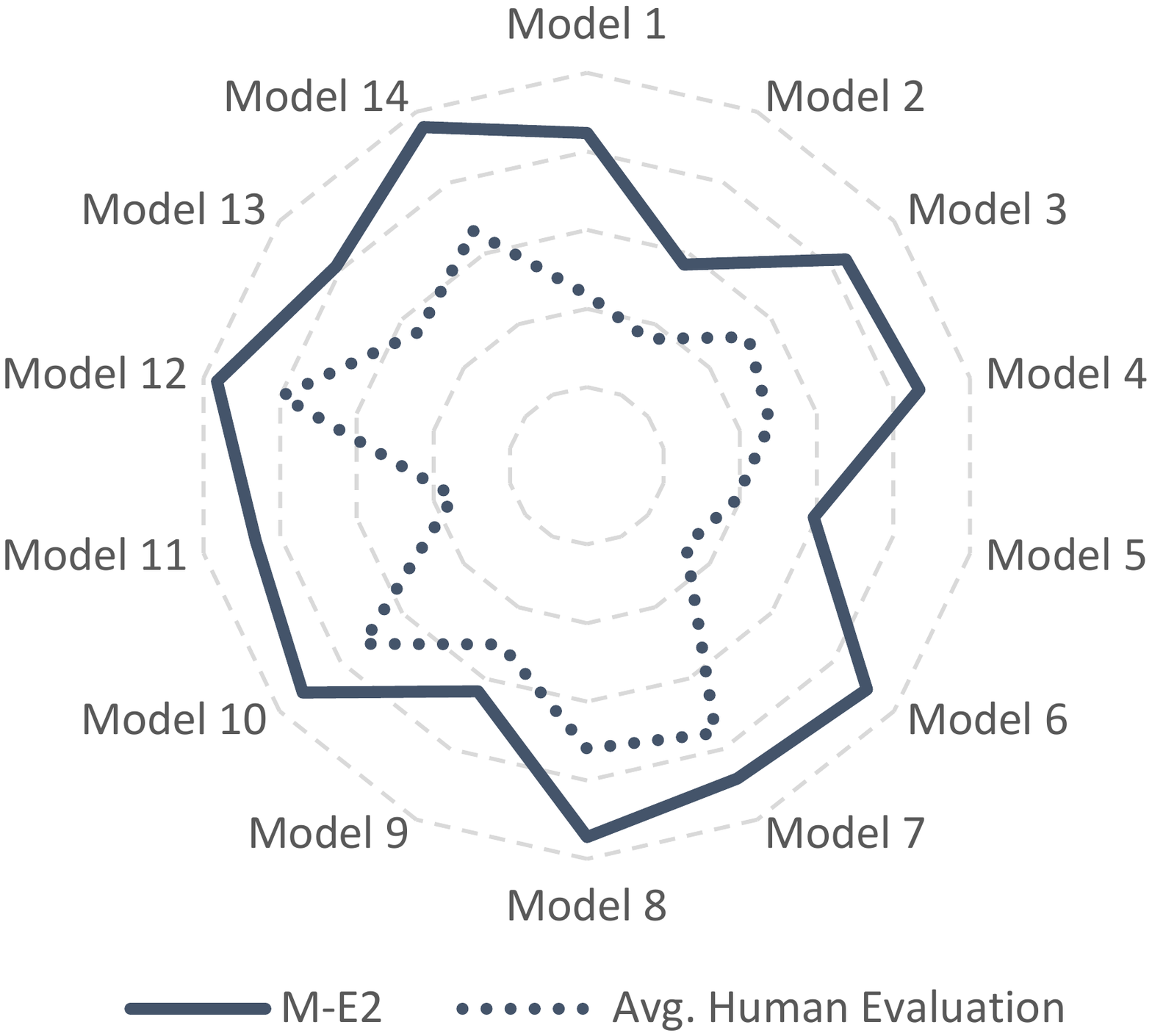}
		\caption{\mAb{} vs. human evaluation. \label{fig:score-trends:mab}}
	\end{subfigure}
	\begin{subfigure}[b]{.49\textwidth}
		\includegraphics[trim={4cm 2cm 4cm 0cm},width=\textwidth]{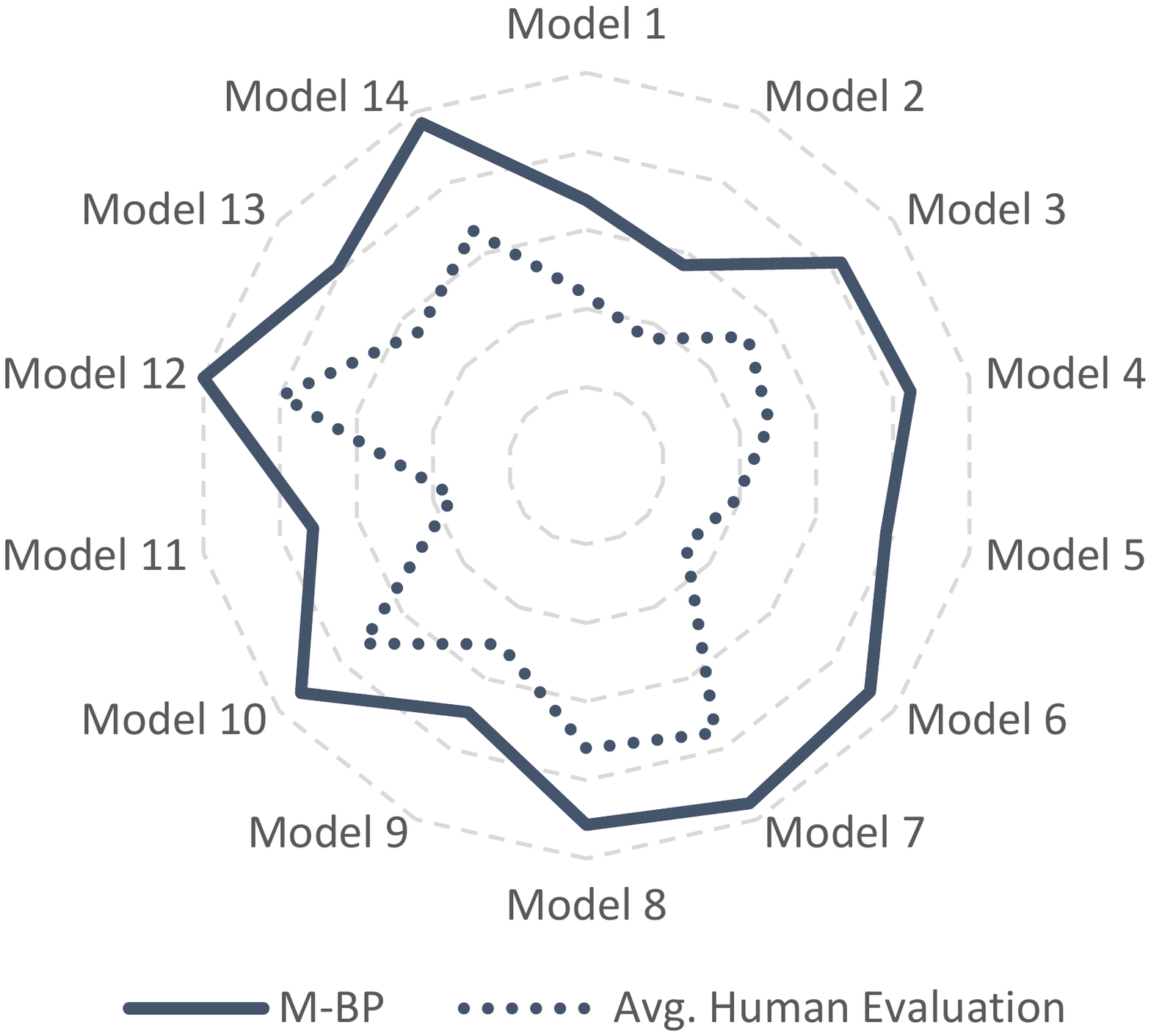}
		\caption{\mB{} vs. human evaluation. \label{fig:score-trends:mb}}
	\end{subfigure}
	\caption{
		Comparison of human assessment and values calculated by our metrics, on all the models of our sample dataset.}
	\label{fig:score-trends}
\end{figure}
Figure~\ref{fig:score-trends} depicts the comparison of the trends of the scores assigned by the three approaches with the average human assessments. The high correlation value of metric \mB{} is reflected in the picture: even though the two curves in Fig.~\ref{fig:score-trends:mb} do not overlap (i.e., the actual scores differ), they describe very similar behavior. Comparable effect is reported in Fig.~\ref{fig:score-trends:mab}, which represents \mAb.
In Fig.~\ref{fig:score-trends:maa}, instead, the curves are touching in some points, however the two shapes are more distinct from each other, thus indicating lower correlation.

From these results we can also infer that, when evaluating the consistency of the flow, human perception tends to give more importance to the position of activities, rather than the actual direction of single edges. This effect entails the ability to abstract from the drawn path of edges and, instead, just focus on the actual flow of the activities.

\paragraph{Limitations}

As Table~\ref{tbl:detailed-scores} reports, for the human evaluation we had, almost for every model, quite high standard deviation values. This is something common when dealing with \emph{intuitive assessment} and we already expected such phenomenon.
Grouping our subjects into clusters, generated according to the scores they share, could help us in improving our evaluation and therefore refining our validation.

\section{Related Work}
\label{sec:relatedwork}

Related to our paper is work on the visual layout of business processes in general and work on the flow of a business process model specifically.

\paragraph{Visual Layout of Business Process Models}

Research on the visual layout of business process models has largely relied on studies done in the field of graph drawing. A considerable body of knowledge exists on how to automatically set the layout of graphical models in order to improve their readability. Studies done in the graph drawing field mainly explored the following visual layout features: edge crossing, edge bends, the minimum angle between edges leaving a node, orthogonality, symmetry, flow direction, edge length variation, and width of layout \cite{old20,old21}. The direct relation between these metrics and understandability was also investigated.

Research on aesthetics of graph layout in general \cite{old21} found that an important feature to users is minimizing line crosses; less important are: minimizing bends, maximizing symmetry; other features were not found to have a significant effect. Research on users' preferences of UML layout/appearance indicates that users rated features as follows: arc crossings, orthogonality, direction of flow, arc bends, text direction, width of layout and font type. Considering process models, \cite{old27} explored understanding of process models by experts and novices in regards to the following layout features: line crossings, edge bends, symmetry, and vicinity of related elements. \cite{old2} investigated user preferences of layout aesthetics for BPMN models, considering heterogeneous user groups with the goal of designing a modeling tool for BPMN. They used line crossings, orthogonal lines, drawing area, line bends, and flow. Findings showed that the aforementioned layout criteria were most relevant for users with average or greater experience and at least basic education in business process modeling. The layout features described above were all identified as part of the findings in our exploratory study.

Another body of work has developed or evaluated algorithms that change an existing layout of a business process model manually or automatically, to match a desirable aesthetical pattern for effective visual layouting of a model. In \cite{old5,old6} both studies present algorithms which are based on a set of constraints targeting a readable layout of a process model (unified flow direction, minimal edge crossing, minimal bendpoints, usage of Manhattan layout). Automatic layout of BPMN models is presented in \cite{old9} and is focused on edge positioning. The study in \cite{old24} presents a comprehensive framework which allows for a personalized process model visualization, meaning that the model's visual appearance can be tailored to the specific needs of different user groups. In addition, in the field of graph drawing, applied research has developed algorithms and related tools to automatically or manually improve visual layout of graphs and thus improve their understandability. GraphEd system in \cite{old8} compared and evaluated different algorithms of graph drawing while considering the following layout criteria: edge length, edge distribution, area, density, bends, crossings, and orthogonal edges. The work presented in \cite{old17} suggested an algorithm which reorders a diagram using orthogonal ordering while preserving the ``mental map'' of the diagram.

The conclusion from the reviewed works is that various attempts to provide precise metrics of specific layout features have been made. Yet, as far as we know, all existing work addresses a conveniently selected set of features, typically those that are immediate to think of and possible to automatically address. In this paper, instead, we present a set of features that are elicited based on human perception. In addition, we extend our own previous work~\cite{Bernstein2015} by focusing on the quantification of the flow direction which has not been addressed comprehensively so far.

\paragraph{Consistency of the Flow}

In \cite{old20}, the consistency of the flow is evaluated with respect to a target direction, which can be parameterized, considering all fragments of all edges.
Specifically, the formalization computes the ratio of fragments pointing towards the target direction divided by the total number of fragments.
This definition is different from the definitions we devised and reported in this paper in two aspects. First, it considers only the edges' angle. Second, in~\cite{old20} authors can deal with just one direction at the time. Therefore, this approach could have problems, for example, with processes containing structures  similar to the ones reported in Fig.~\ref{fig:edge-styles-abstraction-processes}, since it considers each fragment independently; or with the process of Fig.~\ref{fig:motivating-process} which requires the assignment of two directions for each edge.

Related to our work is also the study reported in \cite{old2} which looks into the impact of model aesthetics. The operationalization of flow remains somewhat unclear and is only informally stated as ``\emph{edges are drawn such that they consider the reading direction}''. The applied notion of consistency again focuses on edges, activities as relevant features characterizing flow are not considered.

Another study on the impact of the flow direction of a model is reported in~\cite{Figl2015}. Thereby the paper compares in an experimental setting the effect of different flow directions (either left-to-right, or right-to-left, or top-to-bottom, or bottom-to-top) on model comprehension.
While the models considered in this study all had a clear flow direction, focus of our work is a metric that is able to deal with models that only have a partially consistent flow direction.

\section{Conclusions and Future Work}
\label{sec:conclusions}

The visual layout of the model, the way elements of the model are laid out on the canvas, is an important factor for the user's understanding of the model. Since layout properties are mostly not addressed by modeling languages, and in the absence of enforced layout conventions, the modeler has much freedom to decide on how a model will be laid out. A common terminology in which layout properties can be specified is an essential basis needed for developing an understanding, appropriate conventions, and tools that enforce them.

The contribution of this paper is twofold. First, during a human-centered investigation, visual layout features in business process models were elicited based on what users perceive as important.
Second, we concentrated on one of these features: the consistency of the flow. We operationalized such feature, which is not trivial to quantify, into three possible metrics in a way that closely reflects human perception.
The outcomes of the evaluations we performed, both automatic and involving humans, show that all metrics can be calculated very efficiently. Moreover, we shown that two of the proposed metrics correlate with human perception. In particular, metric \mB{} achieved the highest correlation value. This suggests that, in this context, the position of activities represents a more important features as opposed to the orientation of the edges.

\vspace{1em}
Currently, all the metrics are implemented in a process model tool, allowing to precisely ``measure'' the layout properties of every model.
Yet, validation has been performed just for the flow consistency feature. As future work, we aim to validate additional metrics to corroborate the correlation between the calculated metrics and the human perception of models' layout. In addition, future work may include quantitative studies to experimentally test to what degree these layout features indeed affect user comprehension.

As future work, we also plan to adapt the metrics reported in this paper to use them ``on the fly'', during the process design phase (i.e., even before the process is completely modeled). This way, we could provide continuous and interactive feedbacks to the user modeling the process.


\begin{small}
	\paragraph{Acknowledgements.}
	This work is partially funded by the Austrian Science Fund (FWF) project ``The Modeling Mind: Behavior Patterns in Prozess Modeling'' (P26609).
\end{small}

\bibliographystyle{abbrv}
\bibliography{library-vered,library}

\end{document}